\documentclass[12pt]{iopart}

\newcommand{\eq}[1]{\begin{equation}#1\end{equation}}

\newcommand{\ee}{\mathrm{e}}
\usepackage{iopams}
\usepackage{graphics}
\usepackage{graphicx}

\begin{document}

\title{Entanglement in solvable many-particle models}

\author{Ingo Peschel}
\address{Fachbereich Physik, Freie Universit\"at Berlin, Arnimallee
 14, D-14195 Berlin, Germany}


\begin{abstract}

\noindent Lecture notes for the Brazilian School on Statistical Mechanics \\
  Natal, Brazil, July 18-22, 2011. \\
  The five lectures introduce to the description of entanglement 
in many-particle systems and review the ground-state entanglement features of  
standard solvable lattice models. This is done using a thermodynamic formulation
in which the eigenvalue spectrum of a certain Hamiltonian determines the 
entanglement properties. The methods to obtain it are discussed and results, 
both analytical and numerical, for various cases including time evolution are 
presented. 

\end{abstract}

\vspace{0.8cm}

\noindent {\bf{Preface}}
\vspace{0.2cm}

Entanglement in many-particle quantum states has been a topic of intense research
in recent years with applications in numerics and interesting links to statistical 
physics. It is therefore excellently suited for an advanced course in a summer school. The
following notes correspond closely to five lectures given in July 2011 at the International
Institute of Physics in Natal, Brazil. They are based on a recent review article 
\cite{review09}, but the material has been properly adapted to the purpose. Thus they
contain more introductory examples and certain topics are presented in more detail.
On the other hand, new material from the last two years, as well as supplementary notes
have been added. Throughout the notes, the style is lecture-like with itemized statements.
References are only given in direct connection with the problem at hand, show a preference
of own work and should not be regarded as an exhaustive list. Compared to the version handed 
out in Natal, additional figures have been included and some editing took place.

\vspace{0.3cm}

\noindent {\bf{Contents}} 
\vspace{0.3cm}

\noindent 1. Background and basics \\
2. Free-particle models \\
3. Integrable models \\
4. Entanglement entropies \\
5. Quenches and miscellaneous

\pagebreak

\section{Background and basics}

In this section, I summarize the basic features of entangled states and reduced
density matrices and illustrate them with examples. For further details, see e.g. the short 
review \cite{Ekert/Knight95}.

%
%

\subsection{\bf{Introduction}}

Entanglement is a notion which goes back to 1935 when it was introduced by Schr\"odinger
in a series of three articles (in German, the German word is ``Verschr\"ankung'')
\cite{Schroedinger35}. At the same time Einstein, Podolski and Rosen discussed their famous 
``Gedankenexperiment'', in which they considered two particles with fixed total momentum 
and relative distance. Nowadays this is usually formulated with two spins, and this is also 
where one encounters entanglement first. Entanglement has to do with the features of quantum 
states and the information contained in wave functions. For a long time, it was a topic 
discussed mostly in quantum optics and for systems with few degrees of freedom.

In the last 25 years, however, it has seen a revival with input from very different areas, 
namely
\begin{itemize}
\item the theory of black holes
\item the numerical investigation of quantum chains
\item the field of quantum information
\end{itemize}
In these cases one always deals with large systems and many degrees of freedom.

\noindent In entanglement investigations, one asks the following \emph{question}:
\begin{itemize}
\item given a total system in a certain quantum state $|\Psi\rangle$
\item divide it (in space, or in Hilbert space) in two parts (bipartition)
\item how are the two parts coupled in $|\Psi\rangle$ ?
\end{itemize}
This is more general than looking at, say, a two-point correlation function. And
there is a general way to answer this question, namely one can bring $|\Psi\rangle$
into a standard form, which displays the coupling. This is the {\it{Schmidt 
decomposition}} which we will discuss in a moment. To obtain it in practice, one
uses quantities which determine all properties of a subsystem, namely 
{\it{reduced density matrices}} (RDM’s). They also contain the information on the
entanglement and will be the basic {\it{tool}} throughout the lectures.

The {\it{states}} we will study are the ground states of models which, on the one
hand, are solvable and, on the other hand, have a physical significance, like
tight-binding (hopping) models or spin chains. As in other contexts, they serve
as points of orientation which allow to study the features of the problem
and to develop a feeling and an overall picture. My own interest arose in connection
with the DMRG, where the entanglement turned out to be crucial for the performance
of the method. Entanglement continues to play a role also in other algorithms and their 
design, and in this respect it has quite practical implications. But in these lectures, 
we shall be concerned essentially with the theory.

%

%
%
\subsection{\bf{Schmidt decomposition}}

\begin{figure}[thb]
\centering
\includegraphics[scale=.4]{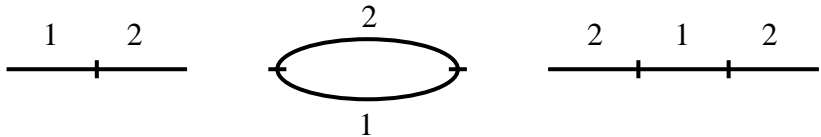}
\caption{Bipartitions: Chain cut in two halves (left), ring cut 
in two halves (centre) and segment in an infinite chain (right).}
\label{fig:geometries}
\end{figure}

Consider a quantum system in state $|\Psi\rangle$ and divide it into two parts 1 and 2.
Then one can write 
\begin{equation}
 |\Psi\rangle = \sum_{m,n} A_{m,n} |\Psi^1_m\rangle  |\Psi^2_n\rangle
\label{state}
\end{equation}
where the $|\Psi^1_m\rangle$ and $|\Psi^2_n\rangle$ are orthonormal bases
in the two Hilbert spaces.\\
Note that one has a double sum and that the matrix $A_{m,n}$ is in general rectangular,
since the dimensions of the Hilbert spaces can differ. 
Nevertheless one can obtain a diagonal form via the so-called singular-value 
decomposition
\begin{equation}
\bf{A= UDV'}
\label{svd}
\end{equation}
where $\bf{U}$ is square and unitary, $\bf{D}$ diagonal and $\bf{V'}$ rectangular with 
orthonormal rows. This gives
\begin{equation}
|\Psi\rangle = \sum_{m,n,k} U_{m,n}D_{n,n}V'_{n,k} |\Psi^1_m\rangle  |\Psi^2_k\rangle 
\label{schmidt1}
\end{equation}
Combining $|\Psi^1_m\rangle$ with $\bf{U}$ and $|\Psi^2_k\rangle$ with $\bf{V'}$
one obtains with $\lambda_{n} = D_{n,n}$ 
\vspace{0.2cm}
\begin{equation}
 |\Psi\rangle = \sum_{n}\lambda_{n}\,|\Phi^1_n\rangle  |\Phi^2_n\rangle
\label{schmidt}
\end{equation}
This is called the Schmidt decomposition (Schmidt 1907) \cite{Schmidt07}. For the history see 
section 1.7. It has the following features
\begin{itemize}
\item Single sum, limited by the smaller Hilbert space
\item New orthonormal sets $|\Phi^{\alpha}_n\rangle$ in both parts
\item $\sum |\lambda_{n}|^2 = 1$ if $|\Psi\rangle$ is normalized
\item Entanglement encoded in the $\lambda_{n}$
\item Limiting cases\\
      $\lambda_{1}=1, \lambda_{n}=0$ for $n>1$: only one term, product state, no entanglement\\
      $\lambda_{n}=\lambda$ for all $n$: all terms equal weight, maximal entanglement  
\end{itemize}
%
This refers to a particular bipartition and one can investigate 
different partitions to obtain a complete picture. Some standard bipartitions
for one-dimensional systems are shown in fig. \ref{fig:geometries}.

%
%
%
%

\subsection{\bf{Examples}}

We give here examples for the Schmidt decomposition in three different systems.
\vspace{0.2cm}

\noindent (a) {\it{Two spins one-half}}
\begin{equation}
 |\Psi_1\rangle = |+\rangle |+\rangle \hspace{1.0cm} \mathrm{product\,\,\, state}
\label{spins1}
\end{equation}
\begin{eqnarray}
 |\Psi_2\rangle &=& a |+\rangle |+\rangle + b |+\rangle |-\rangle  
                    \nonumber \\
                &=& |+\rangle \,\Bigl[ \,a|+\rangle + b |-\rangle \,\Bigr]
\hspace{1.0cm} \mathrm{product\,\,\, state}
\label{spins2}
\end{eqnarray} 
\begin{equation}
 |\Psi_3\rangle = a |+\rangle |+\rangle + d |-\rangle |-\rangle 
 \hspace{1.0cm}\mathrm{entangled \,\,\, state}
\label{spins3}
\end{equation}
All these states are already in Schmidt form. However
\begin{eqnarray}
 |\Psi_4\rangle &=& a |+\rangle |+\rangle + b |+\rangle |-\rangle + 
                      c |-\rangle |+\rangle  \nonumber \\
                &=& |+\rangle \,\,\Bigl[ \,a|+\rangle + b |-\rangle \,\Bigr] +
                      c |-\rangle |+\rangle
\label{spins4}
\end{eqnarray} 
is entangled, but not in Schmidt form, because the two states in subsystem 2 are not 
orthogonal.
\vspace{0.2cm}

\noindent (b) {\it{Two large spins}} \cite{Kaulke/Peschel98}

\begin{figure}[thb]
\centering
\includegraphics[scale=.4]{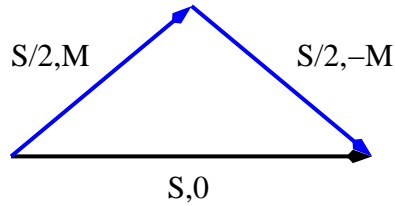}
\caption{The state $|S,0\rangle$ obtained from states in the subsystem.}
\label{fig:clebsch-gordan}
\end{figure}

Consider the ferromagnetic spin one-half Heisenberg chain with $N$ sites
\begin{equation}
 H = -J\,\sum_n \bf{s}_n\bf{s}_{n+1} 
\label{Heisenberg}
\end{equation}
All eigenstates can be written as $|\Psi\rangle = |S,S^z\rangle$ with total spin $S$
and $z$-component $S^z$. In the ground state all spins are parallel, $S=N/2$, and $S^z$ can
be chosen. Choose $S^z=0$ and divide the chain in two halves. Then one can use angular 
momentum addition as illustrated in fig. \ref{fig:clebsch-gordan}\\
%
%
\vspace{0.2cm}

\noindent to obtain
\begin{equation}
|S,0\rangle = \sum_{M=-S/2}^{S/2} c_M \,|S/2,M\rangle_1 \,|S/2,-M\rangle_2
\label{Clebsch1}
\end{equation}
\vspace{0.2cm}
with the Clebsch-Gordan coefficients
\vspace{0.2cm}
\begin{equation}
c_m = \frac {S!}{\sqrt{(2S)!}} \,\frac {S!}{(S/2-M)!(S/2+M)!}
\label{Clebsch2}
\end{equation}
\vspace{0.2cm}
This is the Schmidt form for this state. Its features are
\begin{itemize}
\item Only (S+1) terms, while dimension of Hilbert space is $2^S$
\item For large $S$, $c_M \sim \exp(-2M^2/S)$, Gaussian
\item Analogous formulae for arbitrary $S^z$
\item Special case $S^z=S$, all spins in the $z$-direction. Then $|S,S\rangle$ is 
      a product state
\begin{equation}
|S,S\rangle = |S/2,S/2\rangle_1 |S/2,S/2\rangle_2
\label{Clebsch3}
\end{equation}        
\end{itemize}
\vspace{0.3cm}

\noindent (c) {\it{Two coupled oscillators}} \cite{Han/Kim/Noz99}

Consider the Hamiltonian ($m=\hbar=1$)
\begin{equation}
 H = \frac {1}{2}(p_1^2+\omega_0^2x_1^2) + \frac {1}{2}(p_2^2+\omega_0^2x_2^2)
     + \frac {1}{2} k (x_1-x_2)^2 
\label{two-osc1}
\end{equation}
The eigenfrequencies are $\omega_1^2=\omega_0^2+2k$ and $\omega_2^2=\omega_0^2$
with corresponding normal coordinates
\begin{equation}
y_1=\frac {1}{\sqrt2}(x_1-x_2),\hspace{0.5cm} y_2=\frac {1}{\sqrt2}(x_1+x_2)
\label{normalco}
\end{equation}  
In these coordinates, the ground state is the product of two Gaussians
\begin{equation}
 |\Psi_0\rangle = (\frac{\omega_1\omega_2}{\pi^2})^{1/4} \exp(-\frac {1}{2} [\omega_1 y_1^2+ 
                   \omega_2 y_2^2 ]\,)
\label{two-osc2}
\end{equation}
Then the following formula holds
\vspace{0.2cm}
\begin{equation}
 |\Psi_0\rangle = \sum_{n=0}^{\infty} \frac {(-\mathrm{tanh \,\eta)}^n}{\mathrm{cosh \,\eta}}
                   |\Phi_n(x_1)\rangle |\Phi_n(x_2)\rangle
\label{two-osc3}
\end{equation}
\vspace{0.2cm}
where $\mathrm{exp(4\eta)}=\omega_1/\omega_2$ and the $|\Phi_n\rangle$ are oscillator states 
for a frequency $\bar\omega=\sqrt{\omega_1\omega_2}$ , i.e. in between $\omega_1$ and $\omega_2$.

\vspace{0.2cm}

\noindent \emph{Features}

\begin{itemize}
\item Schmidt states are ``squeezed'' states
\item Coefficients decay exponentially, $\lambda_n^2 \sim \exp(-\varepsilon n)$
\item weak coupling $k$: $\omega_1 \approx \omega_2$ $\rightarrow$ $\eta$ small, $\varepsilon$
      large, rapid decay, weak entanglement
\item strong coupling $k$: $\omega_1 \gg \omega_2$ $\rightarrow$ $\eta$ large, $\varepsilon$
      small, slow decay, strong entanglement 
\end{itemize}
\vspace{0.2cm}
These features are also found for one oscillator in a whole assembly.
%
%

\subsection{\bf{Reduced density matrices}}

The Schmidt structure just discussed can be found from the density
matrices associated with the state $|\Psi\rangle$. This is also the standard
way to obtain it. Starting from the total density matrix
\begin{equation}
 \rho =  |\Psi\rangle \langle\Psi|
\label{rhotot}
\end{equation}
one can, for a chosen division, take the trace over the degrees of freedom in one
part of the system. This gives the reduced density matrix for the other part, i.e.
\begin{equation}
\rho_{1} = \mathrm{tr}_{2}(\rho) \;\;,\;\; \rho_{2} = \mathrm{tr}_{1}(\rho)
\label{rho12}
\end{equation}
These hermitian operators can be used to calculate arbitrary expectation values in
the subsystems. As to the entanglement, assume that $|\Psi\rangle$ has the Schmidt
form (\ref{schmidt}). Then  
\begin{equation} 
 \rho =  |\Psi\rangle \langle\Psi| = \sum_{n,n'}\lambda_n\lambda^*_{n'}
            |\Phi^1_n\rangle  |\Phi^2_n\rangle \langle\Phi^1_{n'}| \langle\Phi^2_{n'}|
\label{rhotot1}
\end{equation}
Taking the traces with the $|\Phi^{\alpha}_n\rangle$ gives $n'=n$ and 
\begin{equation}
 \rho_{\alpha} = \sum_{n}|\lambda_{n}|^2\; |\Phi^{\alpha}_{n}\rangle 
 \langle\Phi^{\alpha}_{n}|\;\;\;,\;\;
 \alpha=1,2
\label{rhodiag}
\end{equation}
This means that
\begin{itemize}
\item $\rho_{1}$ and $\rho_{2}$ have the same non-zero eigenvalues
\item these eigenvalues are given by $w_{n}=|\lambda_{n}|^2$
\item their eigenfunctions are the Schmidt functions $|\Phi^{\alpha}_n\rangle$
\end{itemize}
Therefore the eigenvalue spectrum of the $\rho_{\alpha}$ gives directly the weights
in the Schmidt decomposition and a glance at this spectrum shows the basic entanglement
features of the state, for the chosen bipartition. For this reason, it has also been
termed ``entanglement spectrum'' \cite{Li/Haldane08}. 


\vspace{0.2cm}
\noindent {\it{Remarks}}
\begin{itemize}
\item the $\rho_{\alpha}$ describe mixed states. An expectation value in subsystem 
      $\alpha$  is given by
\begin{equation}
 <A_{\alpha}>\,\, = \sum_{n}|\lambda_{n}|^2\;\langle\Phi^{\alpha}_{n}|A_{\alpha}|\Phi^{\alpha}_{n}
              \rangle 
 \label{mixed}
\end{equation}
\item Since the $\rho_{\alpha}$ are hermitian and have non-negative eigenvalues, one can
      write
\begin{equation}
\rho_{\alpha} = \frac{1}{Z}\; e^{-\mathcal{H}_{\alpha}} \;
\label{rhoexp}
\end{equation}
   where $Z$ is a normalization constant and the operator $\mathcal{H}_{\alpha}$ has been
    termed ``entanglement Hamiltonian''. This form will be encountered permanently in the
    following.
\item The $\rho_{\alpha}$ should not be confused with e.g. the one-particle density matrices,
      which are simple correlation functions.
\end{itemize}
\vspace{0.2cm}
Usually, one starts in a basis where  $|\Psi\rangle$ has the form (\ref{state}). Then
\begin{equation} 
 \rho =  |\Psi\rangle \langle\Psi| = \sum_{m,n,m',n'} A_{m,n} A^*_{m',n'}
            |\Psi^1_m\rangle  |\Psi^2_n\rangle \langle\Psi^1_{m'}| \langle\Psi^2_{n'}|
\label{rhotot2}
\end{equation}
and taking the trace with the  $|\Psi^2_n\rangle$ gives $n'=n$ and
\begin{equation}
 \rho_1 = \sum_{m,m'}\sum_{n} \; A_{m,n} A^{\dagger}_{n,m'} |\Psi^1_{m}\rangle 
 \langle\Psi^1_{m'}|
\label{rhonondiag}
\end{equation}
Thus $\rho_1$ contains the square hermitian matrix {$\bf{AA^{\dagger}}$} and similarly
 $\rho_2$ contains {$\bf{(A^{\dagger}A)^*}$}. The form (\ref{rhodiag}) is then obtained by
diagonalizing these matrices. This is the general approach.
\vspace{0.4cm}

\noindent {\it{Example}}: Two spins one-half \\
A general normalized state is 
\begin{equation}
 |\Psi\rangle = a |+\rangle |+\rangle + b |+\rangle |-\rangle + c |-\rangle |+\rangle + 
                  d|-\rangle |-\rangle
\label{spinsgeneral}
\end{equation}
where $|a|^2+|b|^2+|c|^2+|d|^2=1$. The matrix {$\bf{A}$} is then 
\begin{equation}
\bf{A} =  
    \left(  \begin{array} {ll}
      a & b\\  
      c & d   
  \end{array}  \right)   
  \label{matrixA}
\end{equation}
and one obtains
\begin{equation}
 \bf{AA^{\dagger}} =  
    \left(  \begin{array} {ll}
      aa^*+bb^* & ac^*+bd^*\\  
      ca^*+bd^* & cc^*+dd^*   
  \end{array}  \right)   
  \label{matrixAA}
\end{equation}
Since the trace is one, the eigenvalues are given by
\begin{equation} 
 w_{1,2} = \frac {1}{2} \pm \sqrt{\frac {1}{4}-\mathrm{det(\bf{AA^{\dagger}})}}
\label{wspins}
\end{equation}
The state is entangled if $w_{1,2} \ne 0,1$, i.e. if $\mathrm{det\bf{A}} = ad-bc \ne 0$. 
This includes the state $|\Psi_4\rangle$ in section 1.3, where $a,b,c \ne 0$ and $d=0$.  

%
%
\subsection{\bf{Application: DMRG}}

The density-matrix renormalization group method (DMRG) is a numerical procedure,
which was introduced by Steven White in 1992 \cite{White92,White93} and makes
direct use of the Schmidt decomposition and the reduced density matrices. 
For a review, see \cite{Schollwoeck05}.

\begin{figure}[thb]
\centering
\includegraphics[scale=.4]{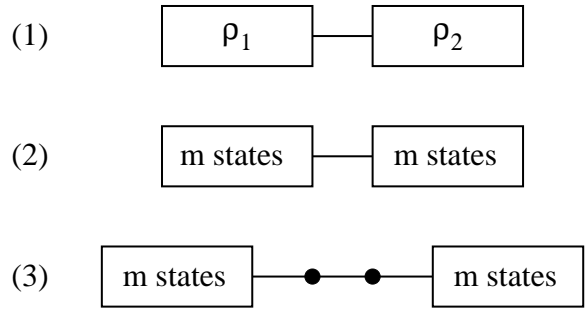}
\caption{Steps in the (infinite-size) DMRG algorithm.}
\label{fig:DMRG}
\end{figure}

Consider a quantum chain, e.g. a spin one-half model, with open ends. Then in the
simplest variant, the following steps take place, compare fig. \ref{fig:DMRG}.

\vspace{0.4cm}

\noindent (0) Start\\ 
Begin with a small system of 5-10 sites.\\
Calculate the ground state {\it{exactly}}.\\

\noindent (1) Schmidt decomposition\\ 
Divide into two halves.\\
Calculate the RDM's.\\
Diagonalize them and obtain the Schmidt coefficients and Schmidt states.\\

\noindent (2) Approximation\\
Keep only the $m$ Schmidt states with largest weights $w_n$.\\
Truncation error: sum of the discarded weights $\sum_{n>m} w_n$. \\

\noindent (3) Enlargement\\
Insert (two) additional sites in the center.\\
Form new Hamiltonian in the basis of kept and additional states.\\
Calculate ground state.\\
Go back to (1) and repeat.\\ 

%
%

For this procedure, the form of the Schmidt spectra is crucial. To have a
good performance, a rapid drop of the $w_n$ is necessary such that only a small number
of Schmidt states has to be kept. In terms of entanglement, the state must be 
{\it{weakly entangled}}. This is satisfied for non-critical chains. For an
Ising model in a transverse field, such a small number as 16 Schmidt states
gives already a fantastic accuracy for the ground-state energy. It is therefore
important to understand the features of RDM spectra and this leads directly
to the study of solvable cases, which is the topic of these lectures.

%
%

\subsection{\bf{Entanglement entropy}}

The full RDM spectra give the clearest impression of the entanglement
in a bipartite system. But it is also desirable to have a simple measure
through one number. Since the eigenvalues of the RDM's can be viewed as
probabilities, one can take the usual entropy, as used in probability theory,
to characterize the $w_n$. This gives the (von Neumannn) entanglement entropy 
\begin{equation}
S_{\alpha}= - \mathrm{tr}(\rho_{\alpha} \ln \rho_{\alpha})= - \sum_n w_n \ln w_n,
\label{ent_def}
\end{equation}
which is the common entanglement measure for bipartitions. It has the following
properties
\begin{itemize}
\item $S_1=S_2 \equiv S$ since the spectra are equal. One can talk of 
      \emph{the} entanglement entropy.
\item $S=0$ for product states.
\item $S$ is maximal if all $w_n$ are equal. \\
      If $w_n= 1/M$ for $n=1,2,\dots,M$ then $S=\ln M$.
\end{itemize}
\par
\noindent The last property leads to a simple interpretation of $S$. Write
\begin{equation}
S=\ln M_{\mathrm{eff}}
\label{ent_inter}
\end{equation}
Then $e^S$ is an effective number of states in the Schmidt decomposition.

A related measure is the R\'enyi entropy 
\begin{equation}
S_{n}= \frac {1}{1-n} \mathrm{ln\,tr}(\rho_{\alpha}^n)
\label{renyi_def}
\end{equation}
where $n$ can also be non-integer. $S_n$ has similar properties as $S$ and the same
extremal values $S=0$ and $S=\mathrm{ln}M$. For $n \rightarrow 1$ write
\begin{equation}
S_{n}= \frac {1}{1-n} \mathrm{ln\,tr}[\rho_{\alpha} \exp((n-1)\ln\rho_{\alpha})]
\label{renyi_limit}
\end{equation}
and expand the exponential function to obtain $S_1=S$. The R\'enyi entropy is somewhat
simpler to calculate, since it contains only a power of $\rho_{\alpha}$. The important point
is that both entropies measure a \emph{mutual connection} and will, in general,
\emph{not} be proportional to the size of a subsystem as usual thermodynamic entropies
are.

%
%
\subsection{\bf{Historical note}}

Erhard Schmidt (1876-1959) obtained his PhD in 1905 with Hilbert in G\"ottingen and
was professor at the Berlin university 1917-1950. He is most widely known by the 
orthogonalization procedure bearing his name. The work linking him
to the quantum problems discussed here, appeared in 1907 in the prestigeous journal 
``Mathematische Annalen'' \cite{Schmidt07}. It was based on his thesis and dealt 
with coupled integral equations with a non-symmetric kernel $K(s,t)$.

In abstract notation, and changing his parameter $\lambda$ to $1/\lambda$, the equations 
were 
\begin{equation}
K\psi = \lambda \phi, \hspace{0.4cm} K'\phi = \lambda \psi
\label{hist1}
\end{equation}
He deduced a spectral representation for $K$
\begin{equation}
K(s,t) = \sum_n \lambda_n \phi_n(s) \psi_n(t)
\label{hist2}
\end{equation}
where $\phi_n$ and $\psi_n$ are the eigenfunctions of the symmetric kernels
$KK'$ and $K'K$ with common eigenvalue $\lambda_n^2$
\begin{equation}
KK'\phi_n = \lambda_n^2 \phi_n, \hspace{0.4cm} K'K\psi_n = \lambda_n^2 \psi_n
\label{hist3}
\end{equation}
One sees that the kernel $K(s,t)$ corresponds to the total wave function, which
for two degrees of freedom is $\Psi(x_1,x_2)$. Moreover, one sees that he already
worked with the quantities which in the present context are called reduced density
matrices. And finally, he discussed best approximations for the kernel based on 
keeping the terms with largest weights, which is the same recipe as used in 
the DMRG. 

The representation of a wave function $\Psi(x_1,x_2)$ in this way was discussed in a 
paper by Schr\"odinger in 1935 \cite{Schroedinger35a}. At that time, unsymmetric kernels 
were already well-known in mathematics, so he referred not to Schmidt but to the textbook 
by Courant and Hilbert. The specialists will notice that the equations (\ref{hist1}),
(\ref{hist3}) with $K=A-B$ and $K'=A+B$ are just the ones appearing in the famous paper 
by Lieb, Schultz and Mattis (1961) \cite{LSM61} where they diagonalize a quadratic form 
in fermions. 

\pagebreak


\section{Free-particle models}

%
%
\subsection{\bf{Solvable cases}}

Before we start the discussion of the free-particle models, which will be the
focus of these lectures, let me list the quantum states for which one can
obtain explicit results for bipartite RDM's and thus for the entanglement
\begin{itemize}
\item Ground states of free-fermion or free-boson systems
\item Ground states of certain integrable models
\item Ground states of conformally invariant models
\item Ground states which have matrix-product form or other simple structures    
\end{itemize}
\par
\noindent An example of the last case was the ferromagnetic ground state in 
section 1.3.

%
%
\subsection{\bf{Free particles, general result}} 

Consider models where the Hamiltonian is a quadratic form in fermionic 
or bosonic operators and defined on a lattice. Two standard examples are
\begin{itemize}
\item Fermionic hopping models with conserved particle number
\begin{equation} 
H=- \frac {1}{2} \sum_{m,n}  t_{m,n} c_m^{\dagger} c_n 
\label{hopping}
\end{equation}
\item Coupled oscillators with eigenfrequency $\omega_0$ 
\begin{equation}
H= \sum_n \left[- \frac {1}{2} \frac {\partial^2}{\partial x^2_n}+
\frac {1}{2} \omega^2_0 x^2_n  \right] + \frac {1}{4} \sum_{m,n} k_{m,n} (x_m- x_n)^2
\label{oscillators}
\end{equation}
\end{itemize}
\vspace{0.2cm} 
For such free-particle models, the reduced density matrices for the ground state
can be written
\begin{equation}
\fbox{$ \displaystyle \hspace{0.5cm}
\rho_{\alpha} = \frac{1}{Z}\; e^{-\mathcal{H}_{\alpha}} \; , \quad
{\mathcal{H}_{\alpha}} = \sum_{l=1}^L \varepsilon_l f_l^{\dagger} f_l \hspace{0.5cm} $}
\label{rhogen}
\end{equation}
Here $L$ is the number of sites in subsystem $\alpha$ and 
the operators $f_l^{\dagger}$, $f_l$ are fermionic or bosonic creation and 
annihilation operators for single-particle states with eigenvalues $\varepsilon_l$.
The $f$'s are related to the original operators in the subsystem by a canonical 
transformation. The constant $Z$ ensures the correct normalization 
$\mathrm{tr}(\rho_{\alpha})=1$.

\noindent Note the following features
\begin{itemize}
\item  $\rho_{\alpha}$ looks thermodynamic. 
\item  the ``entanglement Hamiltonian'' $\mathcal{H}_{\alpha}$ is of the \emph{same type}
        as $H$. 
\end{itemize}
We will see later that  $\mathcal{H}_{\alpha}$ is not the Hamiltonian of the 
subsystem. Therefore (\ref{rhogen}) is not a true Boltzmann formula. Nevertheless, the 
entanglement problem has been reduced to that of a certain Hamiltonian and its 
thermodynamic properties. But first we want to derive the result.
%

%
%
\subsection{\bf{Method 1 - Direct approach}}

The direct method to obtain $\rho_{\alpha}$ is to integrate over the degrees of freedom
outside the subsystem $\alpha$. We illustrate it for the example of two oscillators discussed 
already in section 1.3 \cite{Peschel/Chung99}. The ground state was
\begin{equation}
 |\Psi_0\rangle = (\frac{\omega_1\omega_2}{\pi^2})^{1/4} \exp(-\frac {1}{2} [\omega_1 y_1^2+ 
                   \omega_2 y_2^2 ]\,)
\label{two-osc2}
\end{equation}
One goes through the following steps.

\vspace{0.2cm} 
\noindent \emph{Step I}
\begin{itemize}
\item Write in terms of $x_1$ and $x_2$ 
\item Form $|\Psi_0\rangle \langle\Psi_0|$, i.e. $\Psi_0(x_1,x_2)\Psi_0(x'_1,x'_2)$
\item Set $x'_2=x_2$ and integrate over $x_2$
\item Use $(x_1+x'_1)^2 = 2(x_1^2+x^{'2}_1)-(x_1-x'_1)^2$
\item Result
\end{itemize} 
\begin{equation}
\hspace{-1.3cm}
\rho_1(x_1,x'_1) = C\;\exp{(-\frac{1}{2}(a-b)x_1^2)}\;\exp{(-\frac{b}{4}(x_1-x'_1)^2)}\;
          \exp{(-\frac{1}{2}(a-b)x^{'2}_1)}
\label{rho-osc1}
\end{equation}
where $a=(\omega_1+\omega_2)/2$ and $b=(\omega_1-\omega_2)^2/2(\omega_1+\omega_2)$.

\noindent Due to the derivation, $\rho_1$ has the form of an integral operator. To obtain its
eigenfunctions and eigenvalues, one would have to solve an integral equation.
 
\vspace{0.2cm} 
\noindent \emph{Step II}
\begin{itemize}
\item Determine the differential operator for which (\ref{rho-osc1}) is the $(x_1,x'_1)$ matrix
      element.
\item Observe that
\begin{equation}
\exp{(-\frac{b}{4}(x_1-x'_1)^2)} = 2\,(\frac{\pi}{b})^{1/2}\; 
           \langle x_1|\,\exp{(\frac{1}{b}} \;\frac {\partial^2}{\partial x_1^2}\,)\, |x'_1\rangle
\label{derivative}
\end{equation}
      Proof: Express the operator on the  right in terms of its eigenfunctions 
$\psi_k(x)=(2\pi)^{-1/2} \exp(ikx)$ and integrate over $k$
\item Introduce new coordinates $y^2=bx_1^2/2$ and the frequency $\omega^2/4=(a-b)/b$
\item Result
\end{itemize} 
\begin{equation} 
\rho_1=K\;\exp{(-\frac{1}{4}\omega^2 y^2)}\;\exp{(\frac{1}{2} \;\frac {\partial^2}{\partial y^2})}
         \;\exp{(-\frac{1}{4}\omega^2 y^2)}
\label{rho-osc2}
\end{equation}
If one could simply pull the exponentials together, one would have the Hamiltonian of a 
harmonic oscillator in the exponent. However, the exponentials do not commute.

\vspace{0.2cm} 
\noindent \emph{Step III}
\begin{itemize}
\item Write in terms of boson operators $\alpha,\alpha^{\dagger}$ where
$\alpha = \sqrt{\omega/2}(y+1/\omega \,\partial/\partial y)$
\item Set up equations of motion for Heisenberg operators of $\alpha, \alpha^{\dagger}$ formed 
with $\rho_1$
\item Find Bogoliubov transformation to new boson operators $\beta,\beta^{\dagger}$
\begin{equation} 
\beta=\mathrm{ch}\theta \,\alpha +\mathrm{sh}\theta \,\alpha^{\dagger}, \quad
\beta^{\dagger}=\mathrm{sh}\theta \,\alpha +\mathrm{ch}\theta \,\alpha^{\dagger}   
\label{Bogoljubov}
\end{equation}
such that $\rho_1$ becomes a single exponential. This amounts to another stretching of the 
coordinate $y \rightarrow z$.
\item Result
\end{itemize} 
\begin{equation} 
\rho_1=K\;\exp{(-\varepsilon \;\beta^{\dagger}\beta)}
\label{rho-osc3}
\end{equation}
This is the form announced above. The Hamiltonian $\mathcal{H}_1$ in the exponent describes 
an oscillator with frequency $\varepsilon$ where
\begin{equation} 
\coth(\frac{\varepsilon}{2})=\sqrt{\frac{a}{a-b}}
   = \frac{1}{2}\left[\sqrt{\frac{\omega_1}{\omega_2}}+\sqrt{\frac{\omega_2}{\omega_1}}\right]
\label{eps-osc}
\end{equation}
and its eigenfunctions are those quoted in the Schmidt decomposition (\ref{two-osc3}) when 
expressed in terms of $x_1$.

This derivation can be generalized to any number of oscillators in a larger system, which
proves the general statement for this case. However, one sees that the calculation involves
a number of steps and is already somewhat tedious for the simple case treated above. It is 
therefore fortunate that another much simpler approach exists which we will discuss for 
fermions \cite{Vidal03,Peschel03}.

%
%
\subsection{\bf{Method 2 - Correlation functions}}

Consider a system of free fermions hopping between lattice sites with Hamiltonian
(\ref{hopping}). The ground state is a Slater determinant describing the filled
Fermi sea. In such a state, all many-particle correlation 
functions factorize into products of one-particle functions. For example, 
\begin{equation}
\langle c_m^{\dagger}c_n^{\dagger}c_kc_l\rangle=
\langle c_m^{\dagger}c_l\rangle\langle c_n^{\dagger}c_k\rangle-
\langle c_m^{\dagger}c_k\rangle\langle c_n^{\dagger}c_l\rangle
\label{factor}
\end{equation}
If all sites are in the same subsystem, a calculation using the reduced density matrix
must give the same result. But this is guaranteed by Wick's theorem if $\rho_{\alpha}$
is the exponential of a free-fermion operator 
\begin{equation}
 \rho_{\alpha} = K \exp{(-\sum_{i,j=1}^{L} h_{i,j} c_i^{\dagger} c_j )}
\label{expo2}
\end{equation}
where $i$ and $j$ are sites in the subsystem. Thus $\rho_{\alpha}$ is of the type
given in (\ref{rhogen}). The hopping matrix $h_{i,j}$ is then determined such that 
it gives the correct one-particle correlation functions 
$C_{i,j}=\langle  c_i^{\dagger} c_j\rangle$. This is done in the common diagonal 
representation of both matrices.

If $\phi_l(i)$ are the eigenfunctions of {$\bf{C}$} in the subsystem 
with eigenvalues $\zeta_l$, the transformation
\begin{equation}
 c_i=\sum_l  \phi_l(i) f_l
\label{trans}
\end{equation}
makes the one-particle function diagonal in the new operators $f_l$
\begin{equation}
\langle f_l^{\dagger}f_{l'} \rangle= \zeta_l \,\delta_{l,l'}
\label{diagonalC}
\end{equation}
To obtain this by taking the trace with $\rho_{\alpha}$, the operator
$\mathcal{H}_{\alpha}$ must have the diagonal form given in (\ref{rhogen})
with the two eigenvalues related by
\begin{equation}
 \varepsilon_l=\ln \,(\frac{1-\zeta_l}{\zeta_l}) \hspace{0.5cm} \mathrm{or} \hspace{0.5cm} 
             \zeta_l = \frac{1}{e^{\varepsilon_l}+1}
\label{eigenvalues}
\end{equation}
\par
\noindent \emph{Features}
\begin{itemize}
\item Derivation is very short and clear
\item Valid for any Slater determinant
\item Gaussian nature of the problem, only simplest correlator enters
\item Similar for bosonic case    
\end{itemize}
\par
%

%
%
\subsection{\bf{Example}}

Ring with $N$ sites and nearest-neighbour hopping. The single-particle states are plane
waves and $H$ is diagonalized by putting
\begin{equation}
c_n = \frac{1}{\sqrt{N}} \sum_q \exp{(iqn)} c_q
\label{wave}
\end{equation}
In the ground state, the states are filled up to $q_F$ and the correlation function is 
\begin{eqnarray}
C_{m,n} &=& \frac{1}{N}\sum_q\exp{(-iq(m-n))} \langle  c_q^{\dagger} c_q\rangle \\
       &=& \int_{-q_F}^{q_F} \frac {\mathrm{d}q}{2\pi} \, e^{-iq(m-n)}  ,
                    \quad N \rightarrow \infty\\
       & =& \frac {\sin(q_F(m-n))}{\pi(m-n)}
\label{corrXX}
\end{eqnarray}
Due to the translation invariance, it depends only on the difference $m-n$. For half 
filling $q_F=\pi/2$. Note the oscillation and the power-law decay of the correlations
corresponding to a \emph{critical} system. Mathematically, it is a sort of Hilbert matrix.\\

Choose a segment of $L$ consecutive sites as subsystem, diagonalize the matrix numerically
and order the eigenvalues according to their magnitude. This gives the following fig. 
\ref{fig:spectra_XX}.

\begin{figure}[htb]
\centering
\includegraphics[scale=.6]{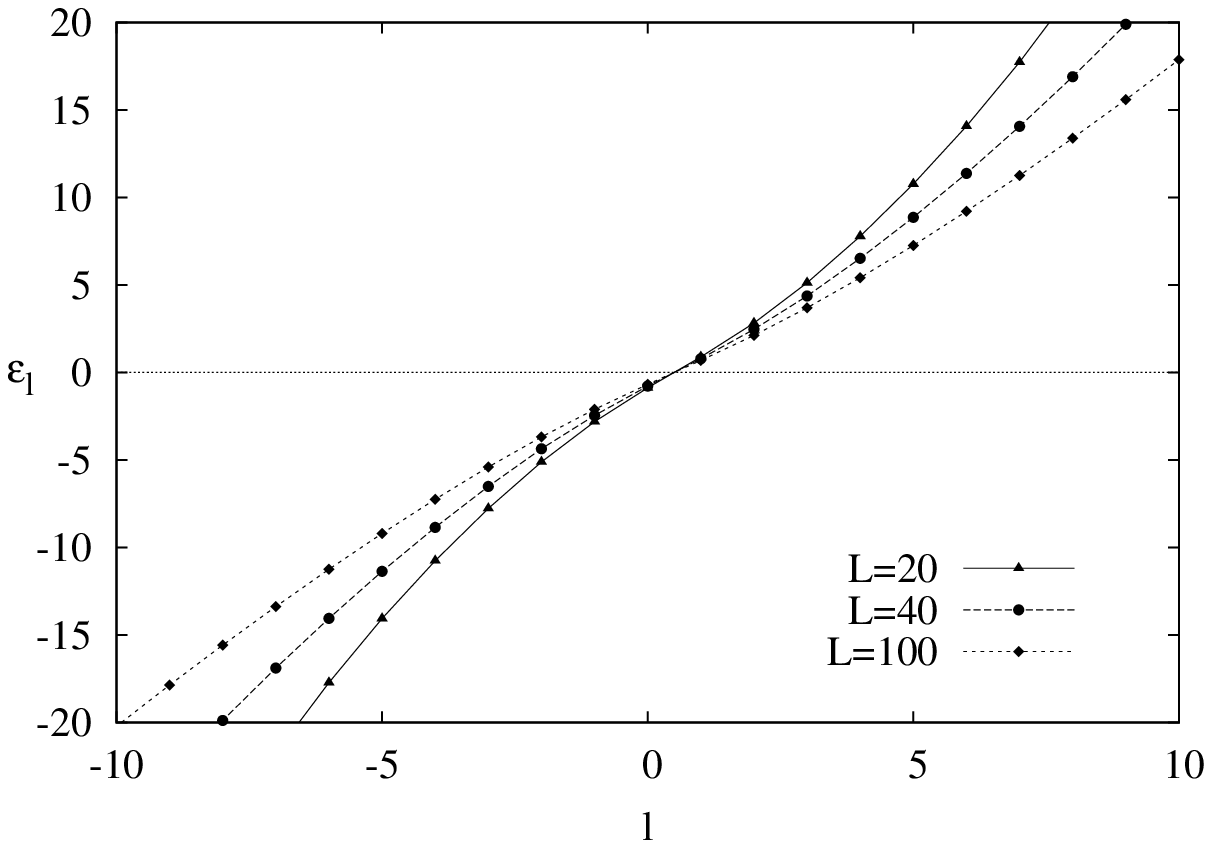}
\includegraphics[scale=.6]{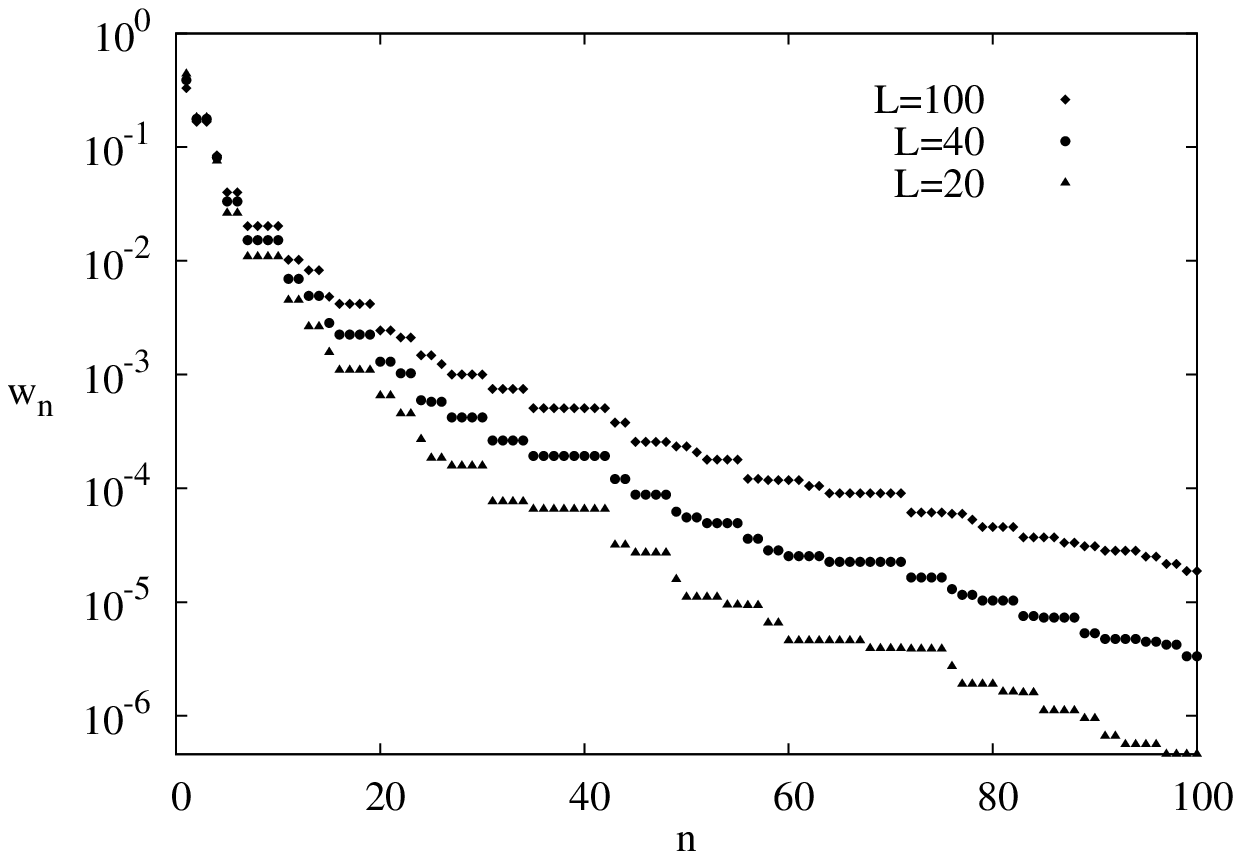}
\caption{Density-matrix spectra for a segment of $L$ sites in an infinite hopping model.
Left: Single-particle eigenvalues $\varepsilon_l$. Right: Total eigenvalues $w_n$.
From \cite{Greifswald08}.\\ Copyright Springer-Verlag, reprinted with permission.}
\label{fig:spectra_XX}
\end{figure}

\pagebreak
 
\noindent \emph{Features}
\begin{itemize}
\item Dispersion of $\varepsilon_l$ roughly linear with curvature
\item Values of order $1$ and larger
\item Curves flatter for larger $L$
\item Rapid initial decrease of the $w_n$
\item Entanglement small, but increasing with $L$
\end{itemize}
\par
%

%
%
\subsection{\bf{Characteristics of the problem}}

\noindent (a) Single-particle eigenfunctions

For the low-lying $\varepsilon_l$, the eigenfunctions are localized near the
boundaries. 

\begin{figure}[thb]
\centering
\includegraphics[scale=.3,angle=270]{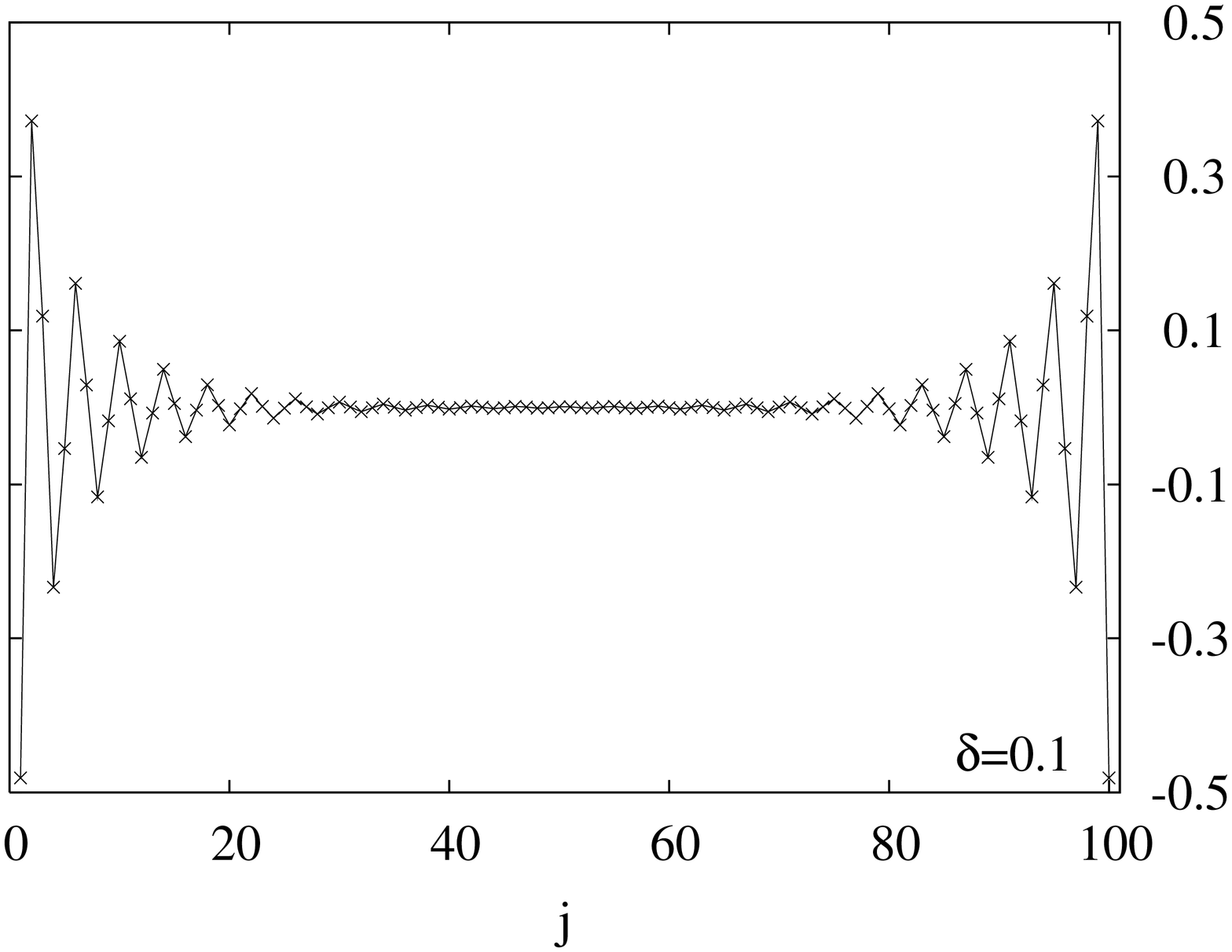}
\includegraphics[scale=.3,angle=270]{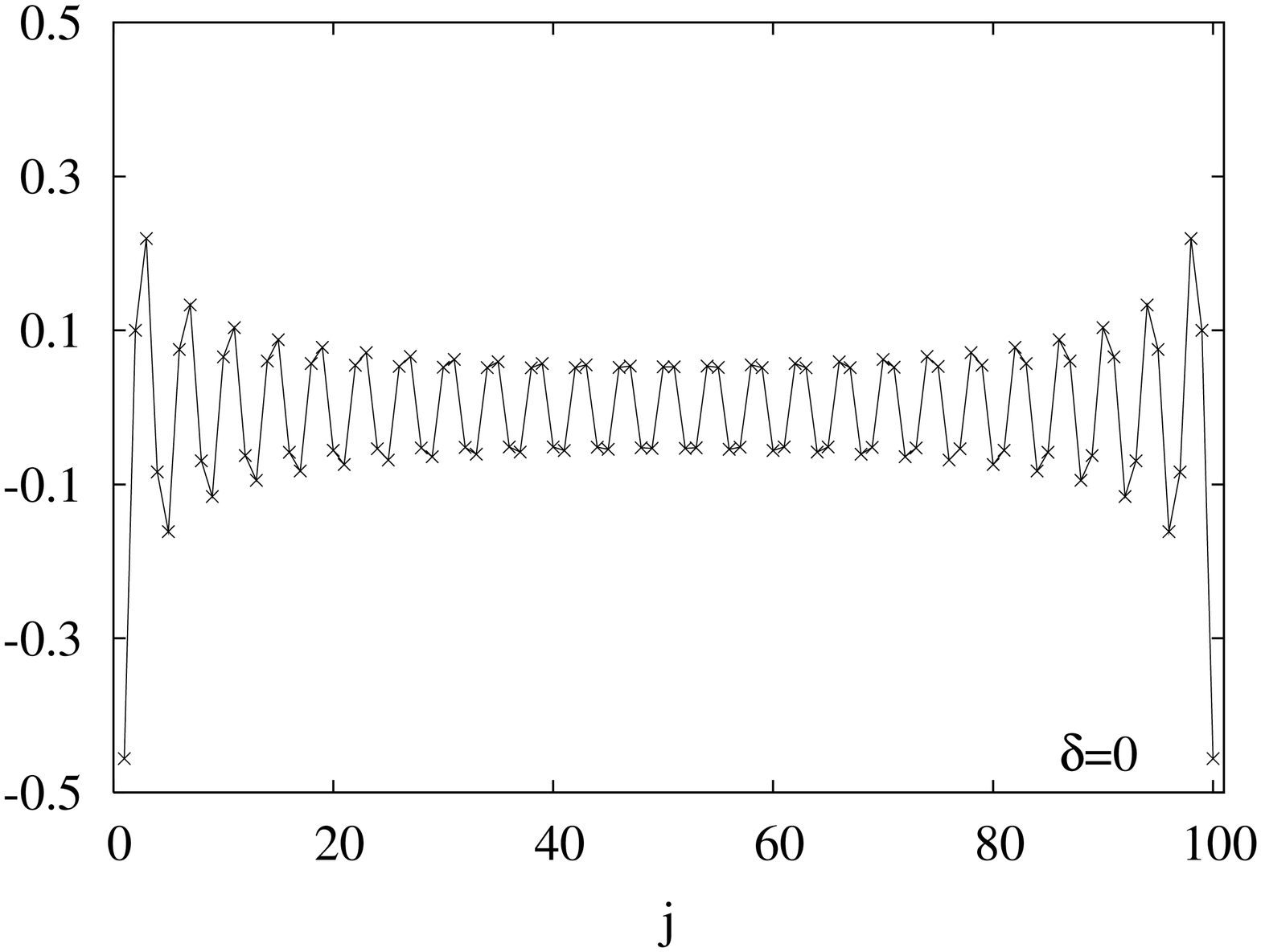}
\caption{Lowest lying single-particle eigenstates in a dimerized ($\delta=0.1$, left) 
and a homogeneous ($\delta=0$, right) hopping model for a segment of $L=100$ sites.
From \cite{Greifswald08}.\\ Copyright Springer-Verlag, reprinted with permission.} 
\label{fig:eigvecs}
\end{figure}

This is shown in fig. \ref{fig:eigvecs} for a non-critical and a critical
hopping chain.
%
%

\noindent \emph{Consequences}
\begin{itemize}
\item Double degeneracy of low $\varepsilon_l$ for segments in non-critical chains
\item Slower decay of the resulting $w_n$
\end{itemize}
\par
\begin{figure}[thb]
\centering
\includegraphics[scale=.4]{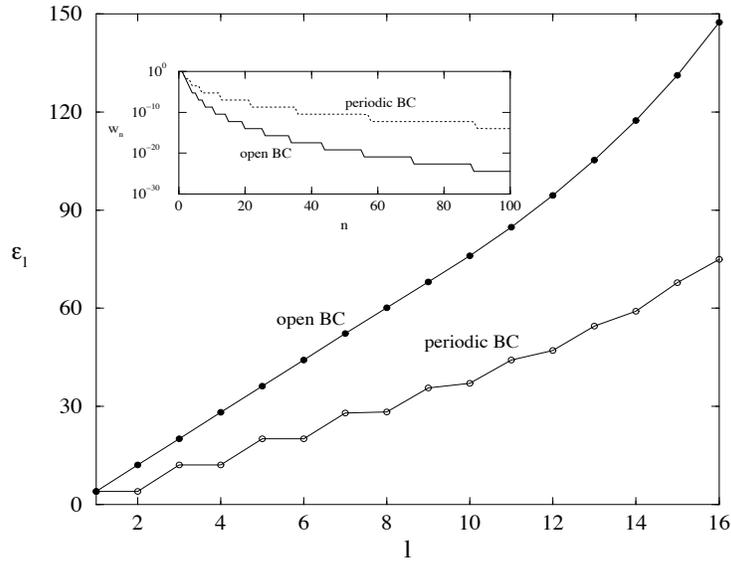}
\caption{Spectra for one-half of an oscillator chain with $k=0.5$ and $L=32$ sites.
         From Chung \cite{Chung02}.}
\label{fig:degeneracy}
\end{figure}

\noindent In fig. \ref{fig:degeneracy} this is illustrated for a half-chain of coupled 
oscillators. Shown are the results both for the open chain (fig. \ref{fig:geometries} left)
where one has one boundary and for the ring (fig. \ref{fig:geometries} centre) where the
subsystem is a segment with two boundaries.

The slower decay of the $w_n$ leads to a poorer performance of the DMRG for rings and explains
why the method is normally used in the open-chain geometry. In two dimensions, whole bands of 
$\varepsilon_l$ arise which are associated with the boundary between the subsystems, see section 4.4. 
In the entanglement entropy this leads to the so-called ``area law''.

\pagebreak

%
%

\vspace{0.4cm}

\noindent (b) Entanglement Hamiltonian $\mathcal{H}_{\alpha}$

In general, this operator is different from the Hamiltonian of the subsystem.

\begin{figure}[htb]
\centering
\includegraphics[scale=.55]{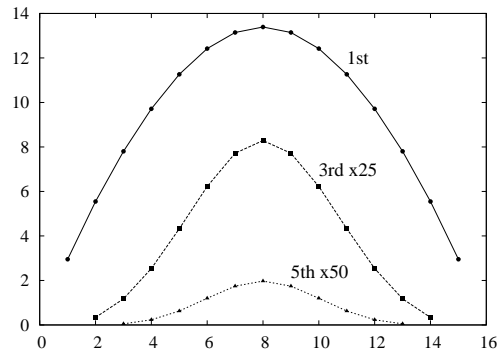}
\caption{Matrix elements in $\mathcal{H}_{\alpha}$ for a hopping model.
 First, third and fifth neighbour hopping in a segment of $L=16$ sites.
 From \cite{review09}. Copyright IOP Publishing. reprinted with permission.} 
\label{fig:heffij}
\end{figure}

This is shown in fig. \ref{fig:heffij} for a segment in a hopping chain. The hopping 
matrix $h_{i,j}$ in (\ref{expo2}) was calculated, using the common eigenfunctions $\phi_l$ of 
$\bf{C}$ and $\bf{h}$, via
%
\begin{equation}
h_{ij} = \sum_l \phi_l(i)\, \varepsilon_l\, \phi_l(j)
\label{matrix_elements}
\end{equation}
\vspace{0.4cm}
%

The dominant elements are those for nearest-neighbour hopping and vary roughly 
parabolically, whereas in the chain they are constant. For a half-chain
one finds half a parabola.

\pagebreak

\noindent (c) Spectrum of {$\bf{C}$}

\begin{figure}[thb]
\centering
\includegraphics[scale=.6]{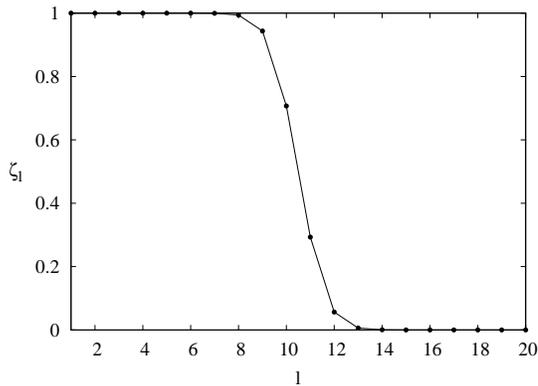}
\caption{Eigenvalues of the correlation matrix for a segment of $L=20$ sites in a hoppping chain.}
\label{fig:zeta}
\end{figure}

In a large subsystem, most of the eigenvalues $\zeta_l$ lie (exponentially) close to 0 and to 1.
\noindent \\
This is illustrated in fig. \ref{fig:zeta} for a segment in a hopping model
%
%

\noindent It can be understood from (52) as follows
\begin{itemize}
\item In the total system, the eigenvalues are $\langle  c_q^{\dagger} c_q\rangle = 0,1$
\item Restricting {$\bf{C}$} to the subsystem changes the spectrum
\item Low-lying states localized near the boundary appear, compare (a)
\item But bulk states remain  
\end{itemize}
In numerics, this leads to the following difficulty. The closeness of $\zeta_l$ to 0 or 1 soon 
exceeds the usual double-precision accuracy. The  $\varepsilon_l$ can then no longer be determined
reliably, unless one works with special techniques. Therefore the values of the $\varepsilon_l$
in most of the figures shown here do not exceed 20-30. However, for the entanglement this does not
matter, since large $\varepsilon_l$ give negligible contributions.

A special role is also played by eigenvalues $\zeta_l=1/2$ corresponding to $\varepsilon_l=0$. 
Such an eigenvalue causes a two-fold degeneracy of all $w_n$ and is therefore seen in the RDM spectrum. 
These zero modes have found much interest recently because they may reflect a symmetry of the real 
Hamiltonian with boundaries.
%
%
\subsection{\bf{Schmidt form for fermions}}

The correlation function approach gives the Schmidt spectra in a very
easy way. But it is also instructive to derive the Schmidt decomposition directly.
This is done in the following way \cite{Klich06}. 
\begin{itemize}

\item Consider a system with $N$ particles. Divide the occupied single-particle states 
$\psi_q(n)$ into the components
\begin{equation}
 \psi_q(n) = \left\{ \begin{array}{r@{\quad:\quad}l}
               \psi_q^1(n) & n \in 1  \\ \psi_q^2(n) & n \in 2
                \end{array} \right.  
\label{sfdivision}
\end{equation}
These are neither orthogonal nor normalized in their subsystems.
 
\item Find new states $\chi_l(n)$ such that their components $\chi_l^1(n)$ and $\chi_l^2(n)$
  are orthogonal in their subsystems. This is done by diagonalizing the overlap matrices
\begin{equation}
M^{\alpha}_{q,q'} = \langle \psi_q^{\alpha}| \psi_{q'}^{\alpha} \rangle, \quad \alpha=1,2
\label{sfoverlap}
\end{equation}
Their eigenvalues are $\zeta_l$ and $1-\zeta_l$ and the new functions 
have the norms
\begin{equation}
\langle \chi_l^1| \chi_l^1 \rangle = \zeta_l, \hspace{0.5cm}
                \langle \chi_l^2| \chi_l^2 \rangle = 1-\zeta_l
\label{sfnorm}
\end{equation}
\item Form normalized states via
\begin{equation}
\phi^1_l = \frac {1}{\sqrt{\zeta_l}}\chi^1_l,\hspace{0.5cm} 
            \phi^2_l = \frac {1}{\sqrt{1-\zeta_l}}\chi^2_l
\label{sfbasis}
\end{equation}
\item Define Fermi operators $a_{\alpha,\,l}$ for the $\phi^{\alpha}_l$.
      Then
\begin{equation}
 |\Psi\rangle = \prod_{l=1}^N \left[\sqrt{\zeta_l} \,a^{\dagger}_{1,\,l} + 
                          \sqrt{1-\zeta_l}\,a^{\dagger}_{2,\,l}\right]|0\rangle
\label{sfschmidt}
\end{equation}
where $|0\rangle$ is the vacuum. This gives the Schmidt decomposition if one 
multiplies out the product.
\end{itemize}
\par
\noindent \emph{Comments}
\begin{itemize}
\item Instead of the $L \times L$ correlation matrix {\bf{C}}, the $N \times N$
      overlap matrix {\bf{M}} appears
\item However, the non-trivial eigenvalues $\zeta_l$ are the same
\item A particle in state $\chi_l$ is found with probability $\zeta_l$ in part 1 and with
      probability $1-\zeta_l$ in part 2
\item If $\zeta_l=0$ the particle is found only in subsystem 2. This \emph{has} to happen, 
      if subsystem 1 cannot accomodate all the $N$ particles.
\item The approach can be applied to continuous systems where $\psi_q(n) \rightarrow 
      \psi_q(x)$ 
\end{itemize}

\noindent The approach shows that the single-particle eigenvalues $\zeta_l$ in one 
subsystem are associated with the eigenvalues $1-\zeta_l$ in the other. The two lead to 
$\pm \varepsilon_l$ and give the same $w_n$-spectrum, as it should be.\\

\noindent \emph{Example} \cite{Calabrese/Mintchev/Vicari11}

\vspace{0.2cm}
\noindent $N$ free fermions on a ring of length $L$, subsystem segment $(-\ell/2,\ell/2)$.

\noindent Single-particle wavefunctions
\begin{equation}
 \psi_q(x) = \frac{1}{\sqrt{L}} \exp{(iqx)}, \quad q=\frac{2\pi}{L}n, 
              \quad n=0,\pm 1,\pm 2, \dots
\label{psi_continuum}
\end{equation}
Overlap matrix in subsystem
\begin{eqnarray}
M^{1}_{q,q'}& =& \langle \psi_q^{1}| \psi_{q'}^{1} \rangle \\
           & =&  \frac {1}{L}\int_{-\ell/2}^{\,\ell/2} dx \, \exp{(-i(q-q')x)}\\
           & =&  \frac {2}{(q-q')L} \sin((q-q')\ell/2)
\label{overlap_continuum}
\end{eqnarray}
Writing $q=2\pi m/L, q'=2\pi n/L $, the matrix becomes
\begin{equation}
M^{1}_{m,n} =  \frac {\sin((\pi\ell/L)(m-n))}{\pi(m-n)}
\label{overlap_continuum1}
\end{equation}
This is the correlation matrix result (\ref{corrXX}) with the substitution 
$q_F \rightarrow \pi\ell/L$. The case $\ell=L/2$ corresponds to half filling,
and one can take over the lattice results for the $\zeta_l$. Choosing a different 
segment of the same length changes $\bf{M}$ but not the eigenvalues.

%
%
%
\subsection{\bf{Some additional details}} 
\begin{itemize}
\item In the correlation function approach, the eigenvalue equation can also be
      written in the form
\begin{equation}
({\bf{1}}-2{\bf{C}})\,\phi_l = \tanh(\frac{\varepsilon_l}{2})\, \phi_l.
\label{purehopping}
\end{equation}
%
\item If the expectation values $F_{i,j}=\langle c_i^{\dagger} c_j^{\dagger}\rangle$ 
and $F^*_{i,j}=\langle c_j c_i\rangle$ are non-zero, they have to be included in 
the considerations. Then for real $F$ the equation becomes
\begin{equation}
(2{\bf{C}}-{\bf{1}}-2{\bf{F}})(2{\bf{C}}-{\bf{1}}+2{\bf{F}})\,\phi_l =
\tanh^2(\frac{\varepsilon_l}{2})\, \phi_l.
\label{pair}
\end{equation}
\item Instead of working with the usual fermions, one can use Majorana fermions
defined by
\begin{equation}
a_{2n-1}=(c_n+c_n^{\dagger}), \hspace{0.5cm} a_{2n}=i(c_n-c_n^{\dagger})
\label{majorana}
\end{equation}
and form the $2L \times 2L$ correlation matrix $\langle  a_m a_n\rangle$ in the 
subsystem. It has eigenvalues  $1 \pm i\tanh(\varepsilon_l/2)$. This is usually 
done if the ``anomalous'' correlation functions $F_{i,j}$ exist.

\item For coupled oscillators, the correlation functions of position variables 
and of momenta, $X_{i,j}=\langle x_i x_j\rangle$ and  $P_{i,j}=\langle p_i p_j\rangle$,
take the place of the Majorana variables. The single-particle eigenvalues then
follow from
\begin{equation}
2 {\bf{P}}\;2 {\bf{X}}\;\phi_l=\coth^2(\frac{\varepsilon_l}{2})\;\phi_l.
\label{boson}
\end{equation}
For the two coupled oscillators treated in section 2.3, one has 
$\langle x_1^2\rangle=(1/\omega_1+1/\omega_2)/4$ and 
$\langle p_1^2\rangle=(\omega_1+\omega_2)/4$ which gives again (\ref{eps-osc}).

\end{itemize}

\pagebreak


\section{Integrable models}

In one dimension one can exploit the relations between quantum spin chains and 
two-dimensional classical models. For \emph{non-critical} integrable models, this allows 
to determine the RDM's and their spectra analytically for \emph{large} systems divided 
in the middle. 
%
%
\subsection{\bf{Transverse Ising model}}

We will discuss the the approach for the Ising model in a transverse field (TI model) with
Hamiltonian
\begin{equation}
H= - \sum_n \sigma^x_n - \lambda \sum_n \sigma^z_n \sigma^z_{n+1},
\label{TI}
\end{equation}
The transverse field has been set to $h=1$. The ground state is non-degenerate for $\lambda < 1$
and asymptotically degenerate with long-range order for $\lambda > 1$.
If rewritten in terms of Fermi operators, $H$ becomes a quadratic form
\begin{equation}
H= - \sum_n (2 \, c^{\dagger}_n c_n - 1) - \lambda \sum_n (c^{\dagger}_n-c_n)(c^{\dagger}_{n+1}+c_{n+1}).
\label{TI2}
\end{equation}
Therefore, according to section 2
\begin{equation}
\rho_{\alpha} = \frac{1}{Z}\; e^{-\mathcal{H}_{\alpha}} \; , \quad
{\mathcal{H}_{\alpha}} = \sum_{l=1}^L \varepsilon_l f_l^{\dagger} f_l 
\label{rhogen1}
\end{equation}
and the $\varepsilon_l$ could be calculated numerically using the correlation functions. 
The present approach will give them analytically.
%
%
\subsection{\bf{Relation to a 2D partition function}}

The TI model has the following features
\begin{itemize}
\item $H$ commutes (up to boundary terms) with a particular (diagonal) transfer matrix $T$
      of an isotropic 2D Ising model on a square lattice
\item Its ground state $|\Psi\rangle$ is the eigenstate of $T$ with maximal eigenvalue
\end{itemize}
From the second property, it follows that one can obtain $|\Psi\rangle$ from an initial state 
$|\Psi_s\rangle$ via
\begin{equation}
|\Psi\rangle \sim \lim_{n \rightarrow \infty} T^n |\Psi_s\rangle
\label{path1}
\end{equation}
\par
\noindent In this way, one has related $|\Psi\rangle$ to the \emph{partition function} of a 
two-dimensional semi-infinite Ising strip. This is a discrete path-integral representation of 
$|\Psi\rangle$. It follows that
\begin{itemize}
\item $\rho=|\Psi\rangle \langle\Psi|$ is given by two such strips

\item $\rho_{\alpha}$ is obtained by tying the two half-strips together
\end{itemize} 

\begin{figure}[htb]
\centering
\includegraphics[scale=.7]{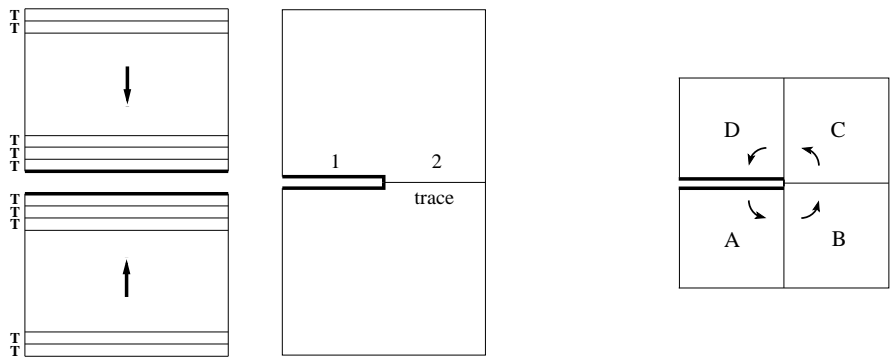}
\caption{Left: Density matrices for a quantum chain as two-dimensional partition
functions. Far left: Expression for $\rho$. Half left: Expression for $\rho_1$.
The matrices are defined by the variables along the thick lines.
Right: Two-dimensional system built from four quadrants with
corresponding corner transfer matrices $A,B,C,D$. The arrows indicate
the direction of transfer. From \cite{Greifswald08}. Copyright Springer-Verlag, 
reprinted with permission.}
\label{fig:pathint}
\end{figure}

\noindent In this way, $\rho_{\alpha}$ is expressed as the partition function of a fully 
infinite strip with a perpendicular cut. This is shown in Fig. \ref{fig:pathint} on the left. 
\par
%
%

%
%
\subsection{\bf{Some transfer matrix formulae}}

Before we discuss the evaluation of this particular partition function, we
list a few relations for conventional Ising transfer matrices.\\

\noindent (a) \emph{One dimension}

Consider the Ising chain with Hamiltonian
\begin{equation}
H= - J \sum_n \sigma_n \sigma_{n+1},
\label{Ising}
\end{equation}
where $\sigma_n = \pm 1$. 
To calculate a partition function, one needs ($K=\beta J$)
\begin{eqnarray}
\exp{(-\beta H)} &=& \exp{(K\sigma_1 \sigma_2)} \,\exp{(K\sigma_2 \sigma_3)}\, 
                     \exp{(K\sigma_3 \sigma_4)} \dots  \nonumber \\
                 &=&  T(\sigma_1,\sigma_2)\,T(\sigma_2,\sigma_3)\,T(\sigma_3,\sigma_4) \dots
\label{transfer1}
\end{eqnarray}
Each $T$ contains the Boltzmann factor for one bond and is a $2 \times 2$
matrix 
\begin{equation}
 T =  
    \left(  \begin{array} {ll}
      e^K & e^{-K}\\  
      e^{-K} & e^K   
  \end{array}  \right)   
  \label{matrixT1}
\end{equation}
Summing over all $\sigma_n = \pm 1 $ multiplies the matrices together and gives, for a ring 
of $N$ sites, the partition function $Z = \mathrm{tr}\,T^N$. In operator form, $T$ can be
written 
\begin{equation}
T = C \,\exp{(K^*\sigma^x)}
  \label{matrixT1op}
\end{equation}
with the so-called dual coupling $K^*$ defined by $\sinh2K^*=1/\sinh2K$. It is large,
if $K$ is small and vice versa.

\vspace{0.4cm}
\noindent (b) \emph{Two dimensions}

\begin{figure}[htb]
\centering
\includegraphics[scale=.55,angle=270]{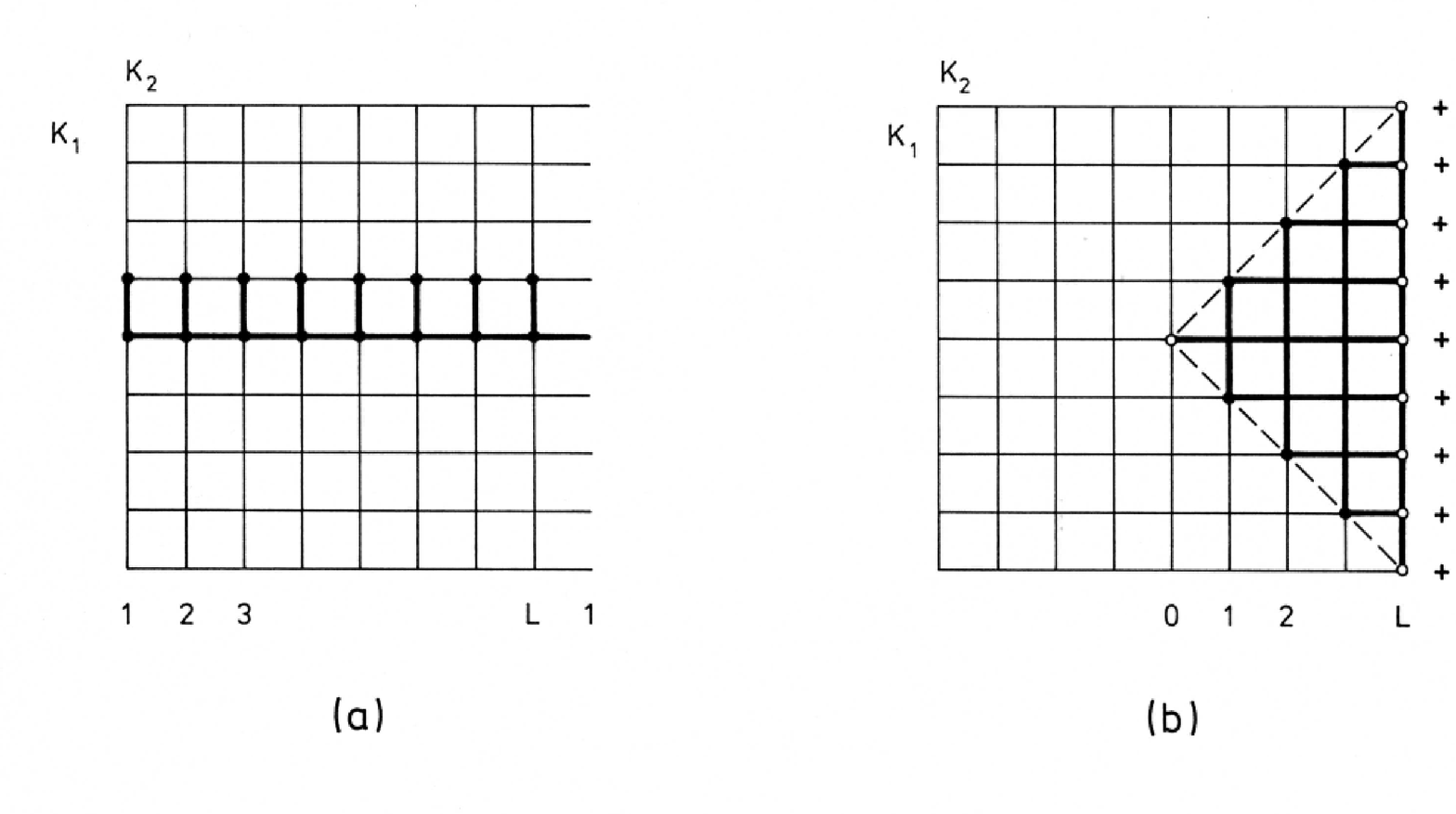}
\caption{Geometry for two types of transfer matrices. (a) Row-to-row transfer matrix
(b) Corner transfer matrix.}
\label{fig:transfer}
\end{figure}

In two dimensions, one can build up a lattice row by row. The transfer matrix then
contains the Boltzmann factors for the vertical and horizontal bonds in one row.
This is shown in fig. \ref{fig:transfer} (a) by the thick lines.

For $N$ sites in a row, $T$ is now a $2^N \times 2^N$ matrix and in operator form
given by
\begin{equation}
T = T_1 T_2 = C^N \,\exp{(K_1^*\sum_n\sigma_n^x)} \, \exp{(K_2 \sum_n\sigma_n^z\sigma_{n+1}^z)}
  \label{matrixT2op}
\end{equation}
where $K_1$ and $K_2$ are the vertical and horizontal couplings, respectively.

\vspace{0.2cm}
\noindent \emph{Features}
\begin{itemize}
\item Terms like in transverse Ising model 
\item However, the exponentials do not commute
\item Exception: $K^*_1, K_2 \ll 1$, strong vertical and weak horizontal bonds. Then
      one can combine the exponentials. This is called the ``Hamiltonian limit''.
\end{itemize}
\par
%

%
%
\subsection{\bf{Corner transfer matrices}}

To calculate the partition function needed for $\rho_{\alpha}$, a kind of ``circular'' 
transfer matrix would be appropriate. This is indicated in fig. \ref{fig:pathint}
on the right. Then $\rho_{\alpha}$ would be given by
\begin{equation} 
\rho_{\alpha} \sim ABCD
  \label{rho-CTM}
\end{equation}
It so happens that such quantities were introduced by Baxter in 1976, see \cite{Baxter82}. 
It turned out that for integrable models they have fascinating and simple properties  
which make them a powerful tool for calulating order parameters.\\

\pagebreak

\noindent (a) \emph{Structure}

In fig. \ref{fig:transfer} (b) a quadrant of a square lattice model is shown. The CTM 
contains all Boltzmann factors indicated by thick lines. The internal variables are summed. 
In operator form, leaving out the prefactor $C$ 
\begin{itemize}
\item Horizontal bonds\\
$\exp{(K_2\,\sigma_0^z\sigma_1^z)}$ \hspace{0.2cm}(1),\hspace{0.3cm}
$\exp{(K_2 \,\sigma_1^z\sigma_2^z)}$ \hspace{0.2cm}(3),\hspace{0.3cm}
$\exp{(K_2 \,\sigma_2^z\sigma_3^z)}$ \hspace{0.2cm}(5), $\dots$
\item Vertical bonds\\
$\exp{(K_1^*\,\sigma_1^x)}$ \hspace{0.2cm}(2),\hspace{0.3cm}
$\exp{(K_1^*\,\sigma_2^x)}$ \hspace{0.2cm}(4),\hspace{0.3cm}
$\exp{(K_1^*\,\sigma_3^x)}$ \hspace{0.2cm}(6), $\dots$
\item All matrices to be multiplied in correct order from bottom to top
\end{itemize}
\par
\noindent Hamiltonian limit
\begin{equation} 
A = e^{-\mathcal{H}_{CTM}}
  \label{CTM1}
\end{equation}
with 
\begin{equation} 
\mathcal{H}_{CTM} = K_1^*\sum_{n \ge 1} 2n \,\sigma_n^x +
                    K_2 \sum_{n \ge 1} (2n-1)\,\sigma_n^z\sigma_{n+1}^z
  \label{H-CTM}
\end{equation}
\noindent \emph{Features}
\begin{itemize}
\item Inhomogeneous TI Hamiltonian 
\item Fields and couplings increase linearly
\item Eigenvalues equidistant for $L \rightarrow \infty$
\begin{equation}
 \varepsilon_l = \left\{ \begin{array}{r@{\quad,\quad}l}
               (2l-1) \varepsilon &   K_1^* <  K_2 \\ 
                2l \varepsilon & K_1^* >  K_2 
                \end{array} \right.  
\label{equidistant}
\end{equation}
\item To be seen directly in the limiting cases
\item Otherwise result of a fermionic calculation
\end{itemize}
\par

\noindent (b) \emph{General case}

So far only the Hamiltonian limit has been considered. The structure of 
$\mathcal{H}_{CTM}$ is then a consequence of the wedge-like geometry. However,
for determining $\rho_{\alpha}$ via (\ref{rho-CTM}), this is not enough, since
in the next quadrant the anisotropy is the other way around.\\

\noindent Amazingly, however, the following holds asymptotically
\begin{itemize}
\item The eigenvalue spectrum of $\mathcal{H}_{CTM}$ has the form (\ref{equidistant})
      for \emph{arbitrary} couplings  
\item In the product $ABCD$, the parameter $\varepsilon$ which gives the level spacing is  
\par
\begin{equation} 
\varepsilon=\pi\,I(k')/I(k),
  \label{epsilon-CTM}
\end{equation}
where $k$ with $0 \le k \le 1$ is either given by $k=\sinh2K_1\sinh2K_2$ or by
$k=1/\sinh2K_1\sinh2K_2$, whichever is smaller than 1. $I(k)$ is the complete elliptic 
integral of the first kind and $k'=\sqrt{1-k^2}$.

\end{itemize}
\par
\begin{figure}[htb]
\centering
\includegraphics[scale=.55]{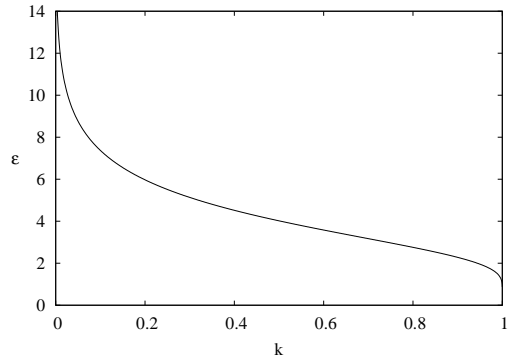}
\caption{Level spacing as a function of the parameter $k$. From \cite{review09}. Copyright IOP 
Publishing, reprinted with permission.}
\label{fig:spacing}
\end{figure}

The parameter $\varepsilon$ diverges for $k \rightarrow 0$ and vanishes for $k \rightarrow 1$,
in both cases logarithmically. It is shown in fig. \ref{fig:spacing}.

%
\noindent The \emph{derivation} uses the integrability of the model, which is contained in the 
so-called star-triangle equations, and a proper elliptic parametrization of the couplings. 
This leads to two parameters, the $k$ appearing above which is connected with the temperature,
and another parameter $u$ which measures the anisotropy, but does not enter the product
$ABCD$. A brief account can be found in the Les Houches lectures of Cardy 1988 \cite{Cardy90}.\\

\noindent (c) \emph{Application to RDM}

\begin{figure}[htb]
\centering
\includegraphics[scale=.33]{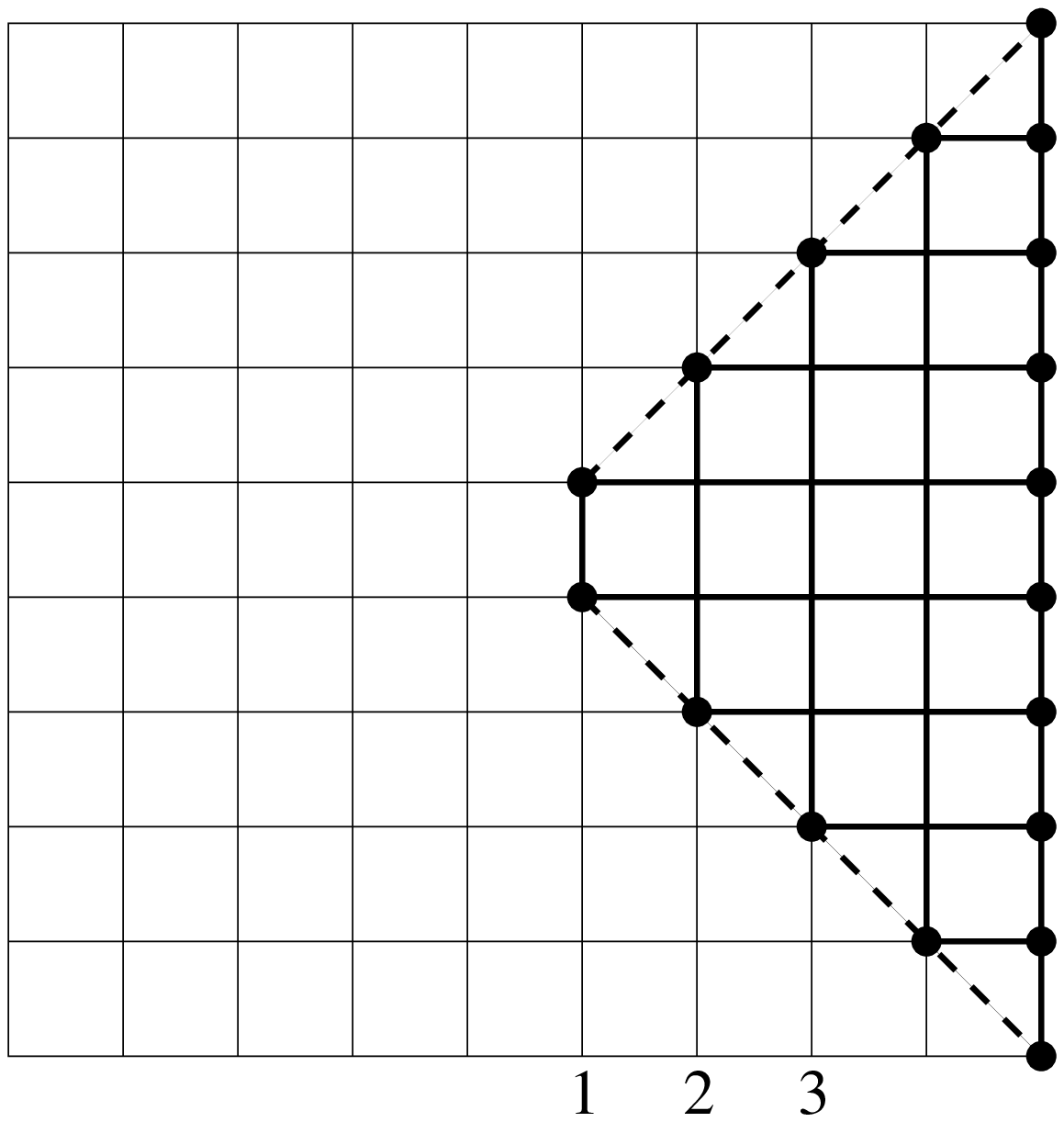}
\hspace{2.5cm}
\includegraphics[scale=.25]{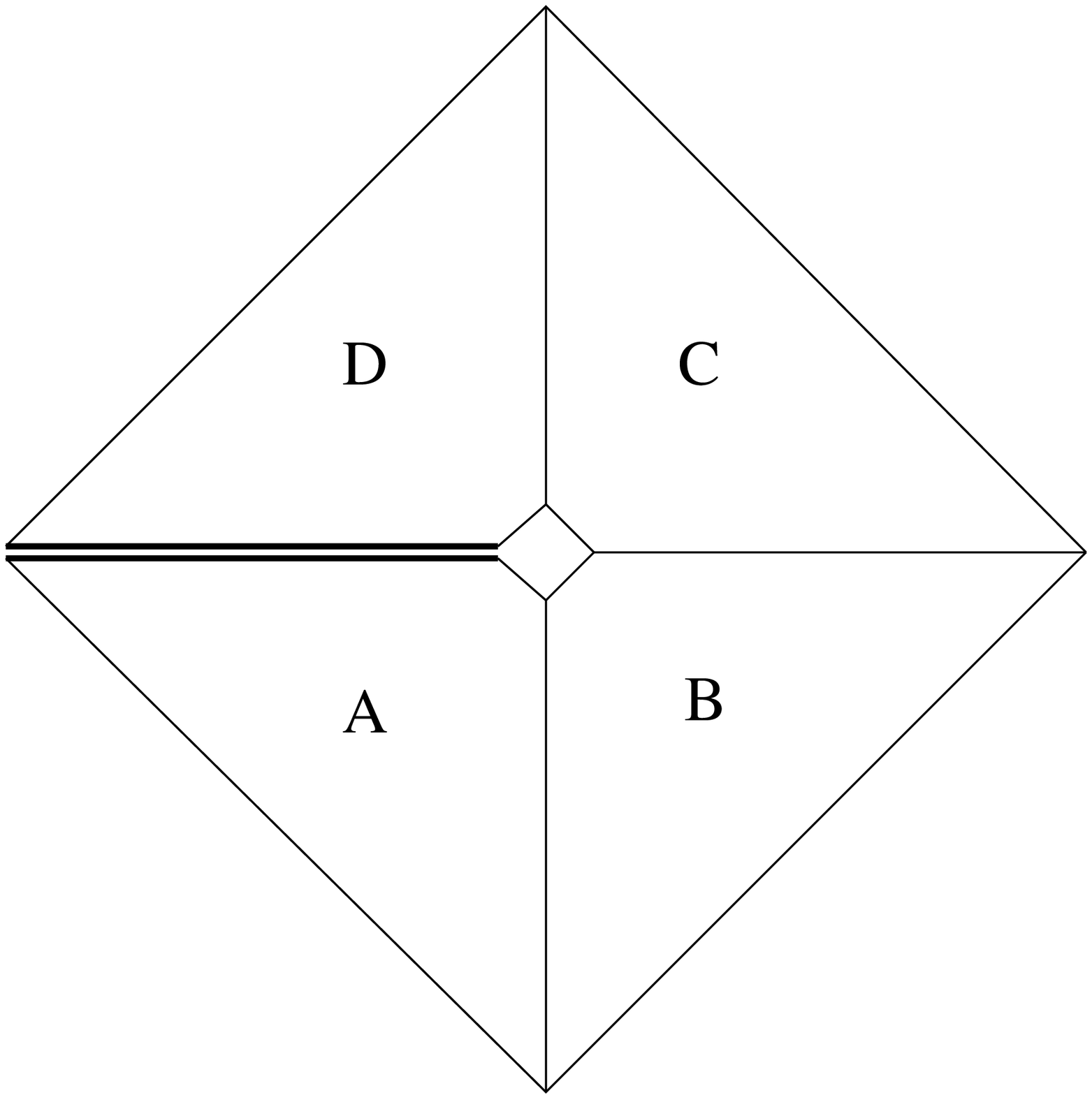}
\caption{Corner transfer matrices without central spin for calculating the RDM. 
Left: single matrix. Right: arrangement of four such matrices giving $\rho_{\alpha}$.}
\label{fig:modCTM}
\end{figure}

The CTM discussed so far can be used for calculating the spontaneous magnetization as expectation
value of the central spin. This is sketched in the supplement. However, this central spin
is an obstacle for the RDM application, because it is common to all four CTM's and prevents 
the division of the system into two parts. To calculate the partition function for $\rho_{\alpha}$, 
one uses the modified CTM shown in fig. \ref{fig:modCTM}. This amounts to an interchange of the 
coefficients $2n$ and $2n-1$ in (\ref{H-CTM}),(\ref{equidistant}).\\
%
%

\vspace{0.4cm}

\noindent Summing up, the result for the single-particle eigenvalues in $\mathcal{H}_{\alpha}$ is
\begin{equation}
 \varepsilon_l= \left\{ \begin{array}{r@{\quad,\quad}r}
(2l+1)\varepsilon & \mathrm{disordered} \; \mathrm{region} \\
2l\varepsilon & \quad \mathrm{ordered} \; \mathrm{region} 
\end{array} \right.  
\label{epsilonCTM}
\end{equation}
where $l=0,1,2,\dots$ and $\varepsilon$ is given by (\ref{epsilon-CTM}). In terms of the 
TI model, the parameter $k$ is 
\begin{equation}
 k = \left\{ \begin{array}{r@{\quad,\quad}r}
\lambda &  \lambda < 1 \\
1/\lambda & \lambda > 1 
\end{array} \right.  
\label{epsilonCTM2}
\end{equation}
%

%
%
\subsection{\bf{Spectra and entanglement}}

\begin{figure}[htb]
\centering
\includegraphics[scale=.43]{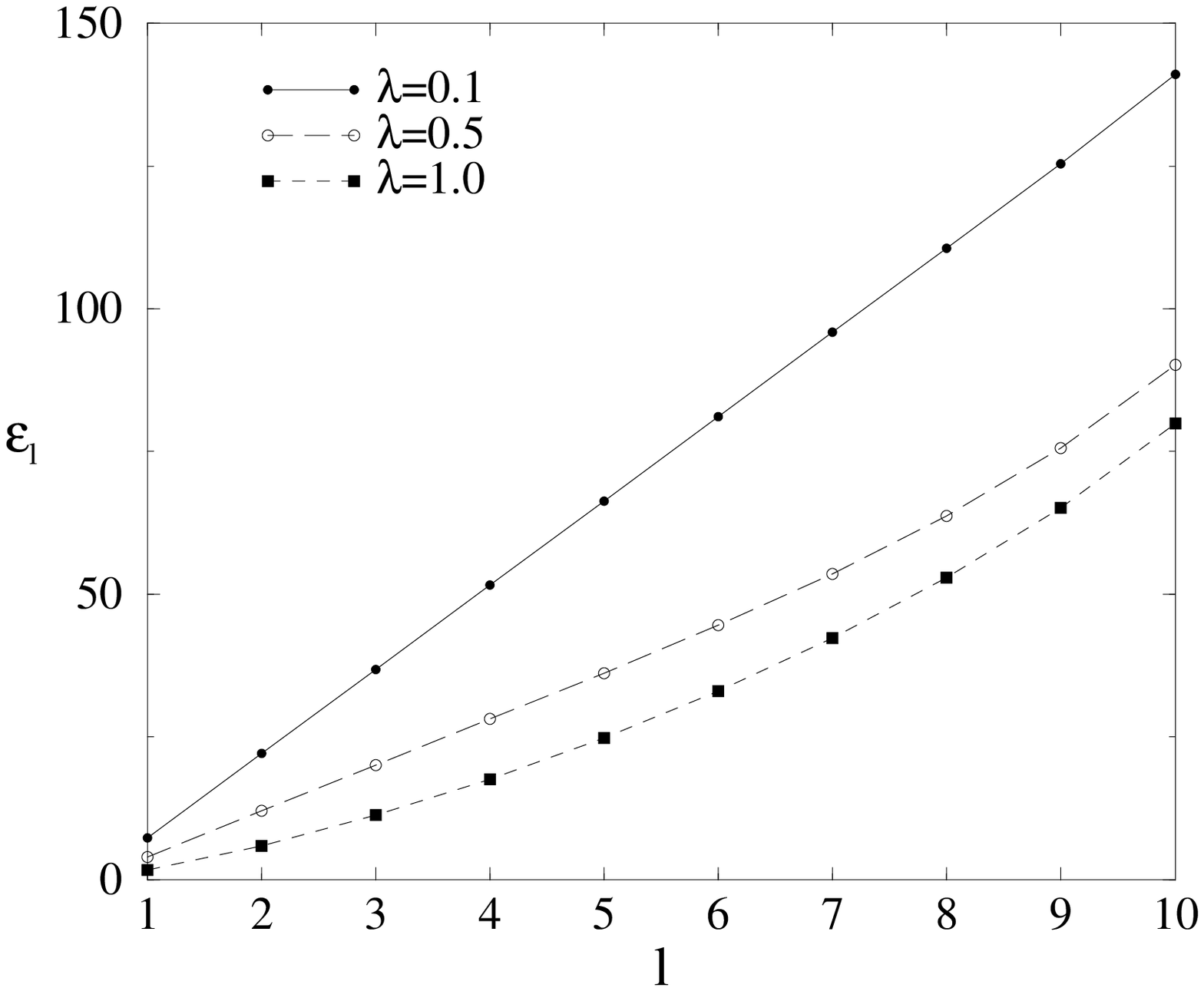}
\includegraphics[scale=.43]{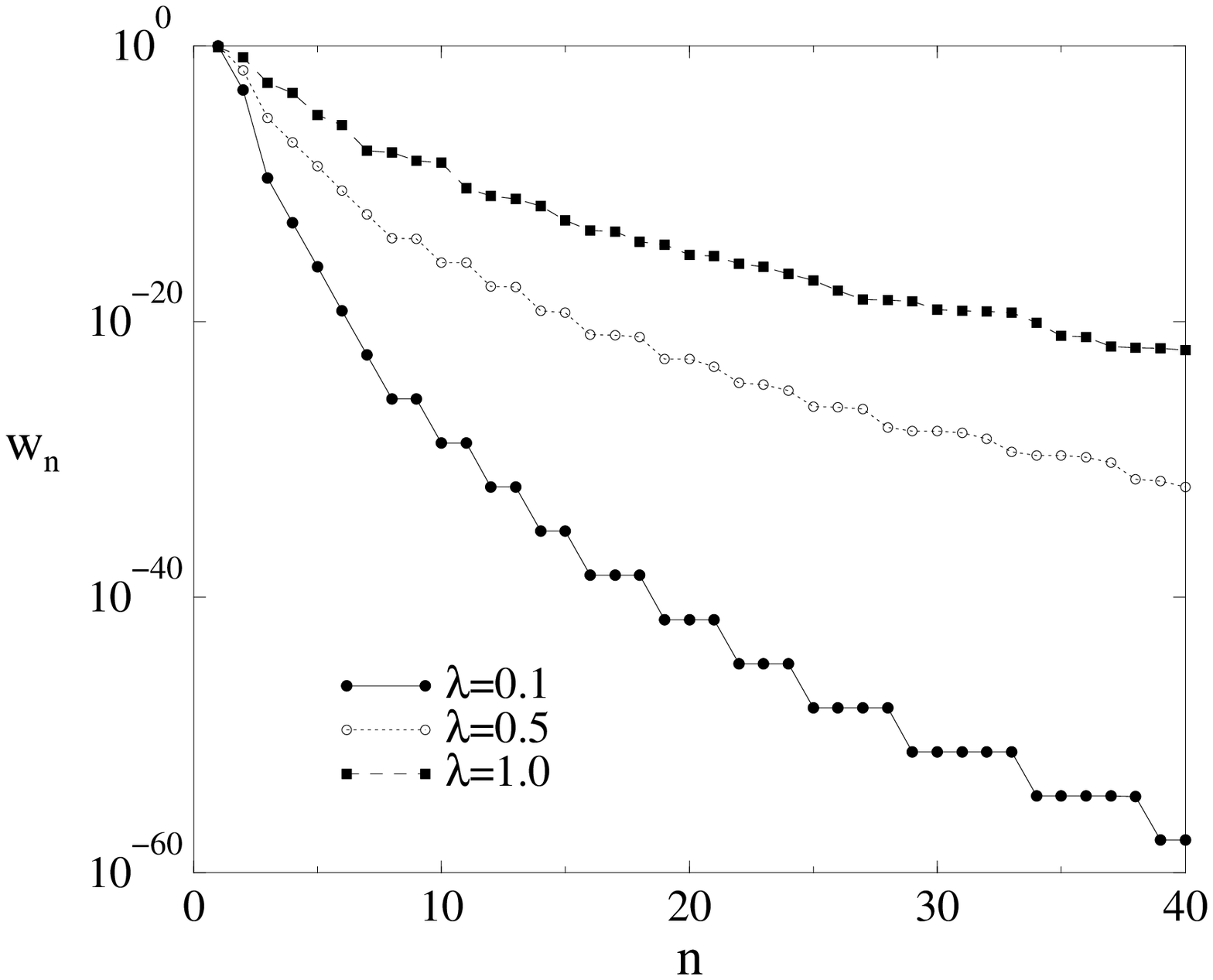}
\caption{Density-matrix spectra for one-half of a transverse Ising chain
with $N=20$ sites in its ground state. Left: All ten single-particle eigenvalues 
$\varepsilon_l$. Right: The largest total eigenvalues $w_n$. From Chung \cite{Chung02}.}
\label{fig:spectra_Ising}
\end{figure}

In Fig. \ref{fig:spectra_Ising}, spectra are shown for a \emph{finite} open TI chain 
with $N=20$ sites, divided in the middle. Thus the subsystem has $L=10$ sites and
there are 10 eigenvalues $\varepsilon_l$. The example displays both the infinite-size 
properties and the modifications by the finite size.

\noindent \emph{Features}
\begin{itemize}
\item Linear behaviour of $\varepsilon_l$ as predicted 
\item Deviations at upper end closer to the critical point $\lambda=1$
\item At $\lambda = 1$ shape as for hopping model
\item $w_n$ decrease extremely rapidly for small $\lambda$ (note the scale)
\item $w_n$-decay slower near criticality, but still impressive
\end{itemize}
\par
%
%
This means that the ground state is weakly entangled. A Schmidt decomposition can be 
truncated safely after about 10 terms and this is the explanation for the fantastic performance 
of the DMRG in this case \cite{Legeza/Fath96}. Note that altogether there are $2^{10}=1024$ 
$w_n$ already in this small system !

\vspace{0.4cm}

\noindent \emph{Behaviour of the $w_n$}
\begin{itemize}
\item Plateaus in $w_n$ for strictly equidistant levels 
\item Behaviour for large $n$ from number of partitions
\begin{equation}
w_n \sim \exp[-a (\ln n)^2]
\label{asymptoticw}
\end{equation}
where $a=\varepsilon \,6/\pi^2$.
\end{itemize}
\par

\noindent The case of a segment cannot be treated by the CTM method, but one can simply include 
the degeneracy seen numerically (section 2.6) into the CTM results. Segments in free-particle
models can be treated by a different method which, however, is more technical and less physical 
than the CTM approach \cite{Its05}.

%
%
\subsection{\bf{Other systems}}

The CTM approach works also for a number of other quantum chains, namely

\begin{itemize}
\item The XY spin chain with Hamiltonian
\begin{equation}
H= - \sum_n \left[ \frac {1+\gamma}{2} \sigma^x_n \sigma^x_{n+1}
+ \frac {1-\gamma}{2} \sigma^y_n \sigma^y_{n+1} \right]  - h \sum_n \sigma^z_n
\label{XY}
\end{equation}
       This generalization of the TI chain also corresponds to a free-fermion problem.\\
       2D problem: Ising model on a triangular lattice
\item The XXZ and XYZ Heisenberg spin chains which contain fermion interactions.\\
       2D problem: Eight-vertex model
\item The oscillator chain with nearest-neighbour coupling.\\
       2D problem: Gaussian model
\end{itemize}
\par
It turns out that the CTM spectrum has the form (\ref{epsilonCTM}) for \emph{all} these
models, even if they contain interactions. Thus one has a \emph{universality} in these 
problems which makes the entanglement properties of all the fermionic systems identical.
Only the parameter $k$ is related differently to the system parameters in each case. For the
oscillator chain, for example, it is given by $k/(1-k)^2=K/\omega_0^2$ if $K$ is the 
nearest-neighbour coupling. This chain is the bosonic analogue of the TI chain, but it has 
no ordered phase. In spite of the different statistics, the $w_n$ spectra are similar and
the asymptotic law (\ref{asymptoticw}) holds with a smaller $a$.

The bosonic formula can also be used to treat exactly a \emph{two-dimensional} lattice of coupled
oscillators which is divided in the middle by a straight line. This is because by making a 
Fourier transformation parallel to the interface, the problem separates into uncoupled chains.
%
\subsection{\bf{Supplement: Onsager formula}}

With the CTM spectra for the 2D Ising model, the famous Onsager formula for the spontaneous
magnetization can be derived in a few lines. Working in the geometry of fig. 
\ref{fig:transfer} (b) and fixing the outer spins as indicated, the expectation value of the
central spin has the form
\begin{equation}
\langle \sigma_0 \rangle = \frac {Z_{+}-Z_{-}} {Z_{+}+Z_{-}}
\label{order1}
\end{equation}
where $Z_{+}$ and $Z_{-}$ are the partition functions with $\sigma_0$ parallel and antiparallel
to the boundary spins, respectively. In terms of the CTM's, this becomes a quotient of traces
\begin{equation}
\langle \sigma_0 \rangle = \frac {\mathrm{tr}(\sigma_0^z \sigma_L^z ABCD)} 
                            {\mathrm{tr}(ABCD)}
\label{order2}
\end{equation}
In the fermionic representation, the operator $\sigma_0^z \sigma_L^z$ can be expressed in terms 
of the operators which diagonalize $\mathcal{H}_{CTM}$ as $\exp{(i\pi\sum_lf^{\dagger}_lf_l)}$. The
trace can then be performed for each $l$ separately and the exponential factor leads to a minus
sign in the numerator. Thus
\begin{equation}
\langle \sigma_0 \rangle = \prod_{l} \frac {1-e^{-\varepsilon_l}} {1+e^{-\varepsilon_l}}                           
\label{order3}
\end{equation}
Since one has to consider the ordered region, one has to choose $\varepsilon_l= (2l-1) \varepsilon$
in (\ref{equidistant}). With $q = e^{-\varepsilon}$ the product then is
\begin{equation}
\langle \sigma_0 \rangle = \prod_{l=1}^{\infty} \frac {1-q^{2l-1}} {1+q^{2l-1}}                           
\label{order4}
\end{equation}
Due to its definition, $q$ is an elliptic nome and the infinite product in (\ref{order4}) has
a simple relation to the elliptic moduli $k$ and $k'$ which appear in $\varepsilon$. This gives
\begin{equation}
\langle \sigma_0 \rangle = (k')^{1/4} = (1-k^2)^{1/8}                           
\label{order5}
\end{equation}
which is Onsager's formula. The parameter $k$ is here $k=1/\sinh2K_1\sinh2K_2$.
It is interesting to note that also Yang in his 1952 proof of Onsager's result \cite{Yang52} 
derived an infinite product equivalent to (\ref{order4}), although his approach was quite 
different. In the CTM formalism, it appears in a natural way, and also the order parameters for 
more complicated models take such product forms, see Baxter's book.

\pagebreak


\section{Entanglement entropies}

We have seen already some RDM spectra, which contain the full entanglement
information. In this section we want to see how their properties translate into
the entanglement entropy. Entanglement entropies are the standard quantities 
considered in this area and have been the topic of a large number of studies. 

%
%
\subsection{\bf{General}}

Due to the form of the $\rho_{\alpha}$, one has the same expressions for the von Neumann entropy 
as in thermodynamics. Thus, $F=U-TS$ with $T=1$, or $S=-F+U$, and the free-particle character of  
$\mathcal{H}_{\alpha}$ gives, as in statistical physics
\begin{equation}
S = \pm \sum_l \ln (1 \pm\mathrm{e}^{-\varepsilon_l})+\sum_l
\frac{\varepsilon_l}{\mathrm{e}^{\varepsilon_l} \pm 1}
\label{ent-thermo}
\end{equation}
where the upper(lower) sign refers to fermions(bosons). From this formula, one
can immediately see some general properties
\begin{itemize}
\item Largest contributions come from small ${\varepsilon_l}$ 
\item Therefore entropy particularly large in critical systems
\item Maximum value for fermions $L \ln 2$ if all $\varepsilon_l=0$
\item If all $\varepsilon_l$ are $m$-fold degenerate, $S$ has $m$ times the value
      without the degeneracy
\end{itemize}

\noindent The last property is an additivity which appears e.g. for uncoupled chains or for
two independent interfaces. 
As to the magnitude, an eigenvalue $\varepsilon_l \sim 1$ also gives a contribution of order 1
to $S$ and the sums converge rapidly for larger $\varepsilon_l$.
%
%
\subsection{\bf{Example: TI chain}}

\begin{figure}[htb]
\centering
\includegraphics[scale=.4]{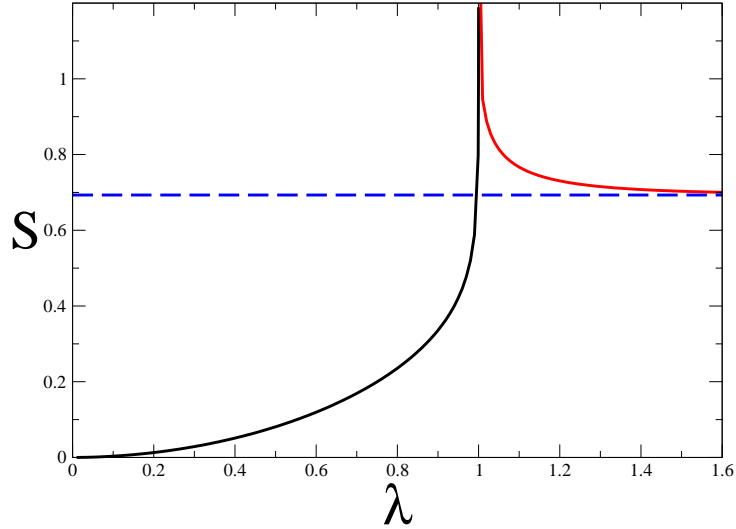}
\caption{Entanglement entropy between the two halves of an infinite TI chain as a function
of $\lambda$. From Calabrese and Cardy \cite{CC04}. Copyright IOP Publishing, reprinted with permission.}
\label{fig:entropyTI}
\end{figure}

 With the spectra found in section 3, it is easy to calculate S for the infinite transverse 
Ising chain. The result of a numerical evaluation is shown in fig. \ref{fig:entropyTI}
%
%
\vspace{0.4cm}

\noindent One notes the following features
\begin{itemize}
\item $S$ vanishes for $\lambda \rightarrow 0$. \\
      Formally: All $\varepsilon_l$ diverge. Physically: $|\Psi\rangle$ becomes product state. 
\item $S$ goes to $\ln 2$ for $\lambda \rightarrow \infty$. \\
      Formally: All $\varepsilon_l$ except one diverge, $\varepsilon_0$ is zero. Physically:
      $|\Psi\rangle$ is superposition of the two product states $|+++ \dots\rangle$ and 
      $|--- \dots\rangle$.
\item $S$ diverges at the critical point $\lambda=1$.
      Formally: $\varepsilon \rightarrow 0$, slope of the dispersion curve goes to zero.
      Physically: State becomes more and more entangled as the correlation length increases.
\end{itemize}

Due to the equidistant single-particle levels, one can even calculate $S$ in closed form.
In the disordered region, one finds with $k=\lambda$
\begin{equation}
   S=  \frac {1} {24} \left[ \;\ln \left( \frac {16} {k^2 k'^2} \right) + (k^2-k'^2)
         \frac {4 I(k) I(k')} {\pi} \right] ,
   \label{eqn:STI1}
\end{equation}
A similar expression with an additional contribution of $\ln 2$ coming from the 
eigenvalue $\varepsilon_0 = 0$ holds in the ordered region. \\
From this, one can extract
the behaviour near $k=1$
\begin{equation}
   S=  \frac {1} {12} \;\ln \left( \frac {8} {1-k} \right)
   \label{eqn:STI2}
\end{equation}
and since the correlation length is given by $\xi \sim 1/(1-k)$, this can be written
\begin{equation}
   S=  \frac {1} {12} \;\ln \xi
   \label{eqn:STI3}
\end{equation}

\noindent which shows a \emph{logarithmic} critical behaviour. The effective number of states in the
Schmidt decomposition, however, has normal power-law behaviour
\begin{equation}
  M_{\mathrm{eff}}  \sim  \xi^{1/12}
   \label{eqn:MeffTI}
\end{equation}
In this sense, the coefficient of the logarithm is a critical exponent.

\noindent The R\'enyi entropies are 
\begin{equation}
S_n = \frac {1}{1-n} \sum_l \ln \frac{(1+e^{-n\varepsilon_l})}{(1+e^{-\varepsilon_l})^n}
   \label{eqn:renyiTI1}
\end{equation}
and lead to more complicated closed expressions, but the critical behaviour is analogous
\begin{equation}
S_n =  \frac {1} {24} (1+\frac{1}{n})\;\ln \xi
   \label{eqn:renyiTI2}
\end{equation}
An unusual structure is seen if one looks at the next (subleading) terms in the expansion.
One finds that they are of the form $\xi^{-k/n}$ with $k=1,2,3\dots$, i.e. the powers depend
on the R\'enyi index $n$ which determines the number of windings in the path integral 
for $\rho_{\alpha}^n$ \cite{CCP10,Ercolessi11}. The same phenomenon is encountered for order 
parameters on such Riemann manifolds.
%
%
\subsection{\bf{Critical chains}}

At a critical point, one has to work with finite subsystems. The spectra for a hopping model 
have already been shown in section 2.5, and a marked size dependence was noted. The dispersion
curves of the $\varepsilon_l$ became flatter with increasing $L$. This gives an increase of $S$.
From (\ref{eqn:STI3}) one can already guess that $\xi$ will be replaced by the length of the
subsystem, and this is in fact the case. The asymptotic formula is
\vspace{0.2cm}
\begin{equation}
 S = \nu \frac {c}{6} \, \ln L + k
\label{eq:Scrit3}
\end{equation}
\vspace{0.2cm}

\noindent \emph{Features}
\begin{itemize}
\item $\nu=1,2$ number of contact points between subsystem and the rest
\item $k$ non-universal constant (subleading term)
\item $c$ central charge, from conformal considerations,
      $c=1/2$ for TI model, $c=1$ for hopping model
\end{itemize}

This result can be understood for the hopping model as follows. The $\varepsilon_l$ curves for 
small systems are not linear, but show curvature. However, for large $L$, more precisely for
large $\ln L$, one can use a continuum approximation to the eigenvalue equation to derive the 
formula, for a segment in a chain, 
\begin{equation}
 \varepsilon_l = \pm \;\frac{\pi^2}{2\ln L} (2l-1) \;, \;\;\; l = 1,2,3\dots 
\label{epsilonconf}
\end{equation}
Using this in (\ref{ent-thermo}) and changing the sums into integrals gives
\begin{eqnarray}
  S =  \frac {2 \,\ln L}{\pi^2} \, \left [ \, \int_{0}^{\infty} \mathrm{d}\varepsilon \; 
       \ln(1+ \exp(-\varepsilon)) +
       \int_{0}^{\infty} \mathrm{d}\varepsilon \;\frac {\varepsilon}{\exp(\varepsilon)+1} \right ]
\label{eqn:Scrit1}
\end{eqnarray}
and since both integrals equal $\pi^2/12$ one finds  
\begin{equation}
 S = \frac {1}{3} \, \ln L
\label{eq:Scrit2}
\end{equation}
In numerical calculations, this logarithmic law can be seen already in relatively small systems,
where (\ref{epsilonconf}) does not yet hold, but one has approximately 
$\ln L \rightarrow \ln L +2.5$ for the first eigenvalues.

The expression for the R\'enyi entropy follows in the same way by going over to integrals in
(\ref{eqn:renyiTI1}) and gives for a segment
\begin{equation}
 S_n = \frac {1}{6} (1+\frac{1}{n})\;\ln L
\label{eq:renyicrit}
\end{equation}
%
%
%
\subsection{\bf{Higher dimensions}}

\begin{figure}[htb]
\centering
\includegraphics[scale=.45]{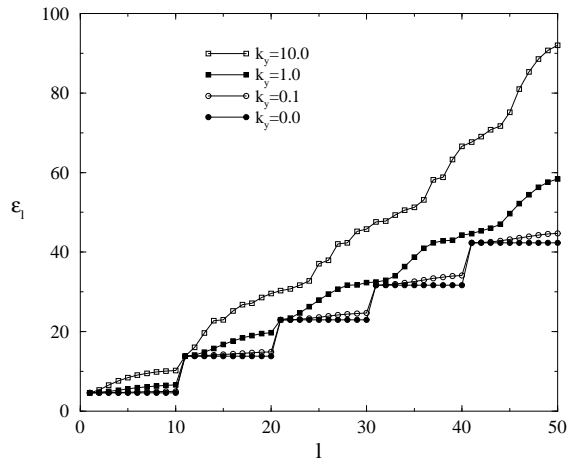}
\caption{Single-particle eigenvalues for one-half of a $10 \times 10$ system of coupled
oscillators with $\omega_0=k_x=1$ and different couplings $k_y$. From \cite{Chung/Peschel00}.
Copyright APS, reprinted with permission.}
\label{fig:eps_osc2d}
\end{figure}

As mentioned in section 2.6, one finds bands of $\varepsilon_l$ in two dimensions. This is 
illustrated in fig. \ref{fig:eps_osc2d} for a $10 \times 10$ square lattice of
coupled oscillators, divided into two halves. The vertical coupling was varied and one 
can see how the plateaus with 10 levels (for 10 uncoupled chains) develop into bands. The states
in a band can be indexed by a vertical momentum $q_y=q$.\\
%
%

\noindent For the entropy, this has the following consequences
\begin{itemize}
\item Without coupling: each chain gives the same contribution $s$ to the total
      entanglement entropy. Thus for $M$ chains one has $S=M\,s$.
\item With coupling: one has to add up the contributions $s(q)$ for each value
      of $q$. For large $M$
\begin{equation}
  S=  \sum_{q} s(q) \simeq  M \int_0^{\pi} \frac {\mathrm{d}q}{\pi} s(q) 
   \label{eqn:Sband}
\end{equation}
\item Therefore $S$ proportional to the length of the interface between the subsystems.
\item In three dimensions: area of the interface
\item Also for other geometries
\end{itemize}

This is the so-called \emph{area law} for the entanglement entropy. For fermionic
critical systems, however, one has logarithmic corrections. For a system with typical 
size $L$ in $d$ dimensions, one finds
\begin{equation}
S \sim L^{d-1}\ln L
   \label{eqn:Sferm}
\end{equation}
if the the state corresponds to a finite Fermi surface. This can be proved exactly by
putting bounds on $S$ \cite{Wolf06,Gioev/Klich06}, see section 4.7.

%
%
\subsection{\bf{Entanglement across a defect}}

\begin{figure}[htb]
\centering
\includegraphics[scale=.5]{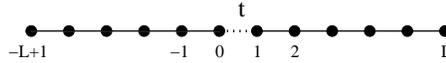}
\caption{Transverse Ising chain with a bond defect.}
\label{fig:bond_defect}
\end{figure}

Since the entanglement is a kind of boundary phenomenon, one expects that in will be
changed by a modification of the interface between the subsystems. This has been 
investigated for hopping chains and critical TI chains with a modified bond, as shown in 
fig. \ref{fig:bond_defect}.

\pagebreak
%
%

\begin{figure}
\center
\includegraphics[scale=0.7]{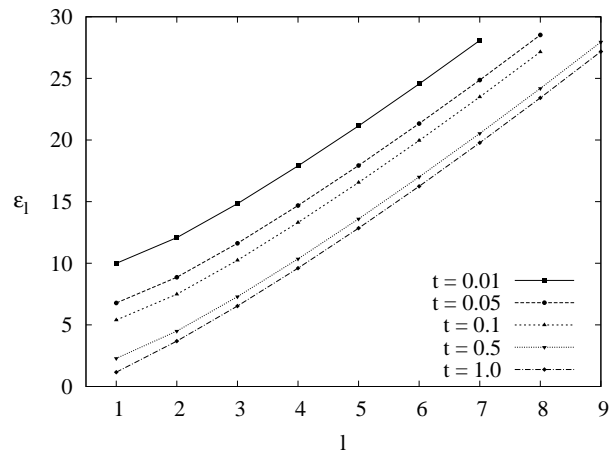}
\caption{Single-particle eigenvalues $\varepsilon_l$ as a function of the defect
strength for TI chains with $2L=300$ sites. From \cite{Eisler/Peschel10}. Copyright Wiley-VCH,
reprinted with permisson.}
\label{fig:defect_spectra}
\end{figure}

\noindent \emph{Limiting cases}
\begin{itemize}
\item Chain cut by defect, $t=0$: no entanglement, $S=0$
\item Chain homogeneous, $t=1$: logarithmic law (\ref{eq:Scrit3}), $S \sim \ln L$  
\end{itemize}
What happens in between ? Numerical results for the $\varepsilon_l$
are shown in fig. \ref{fig:defect_spectra}.
%
%
\vspace{0.2cm}

\noindent \emph{Features}
\begin{itemize}
\item Development of a gap at the lower end of the spectrum 
\item Upward shift of the whole dispersion curve as $t$ goes to zero
\item Therefore decrease of $S$ for fixed $L$
\item Logarithmic law for $S$ remains valid
\item But $c \rightarrow c_\mathrm{eff}(t)$.
\end{itemize}

\begin{figure}
\center
\includegraphics[scale=0.5]{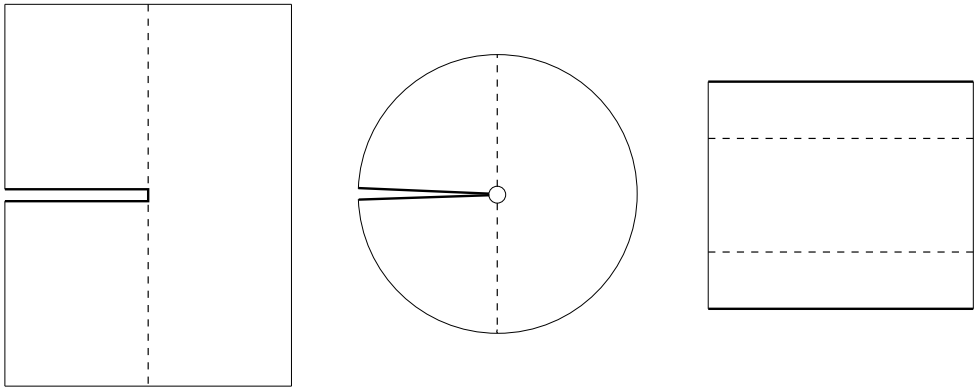}
\caption{Representation of $\rho_{\alpha}$ for a chain with a defect in the centre by 
two-dimensional partition functions. Left: Original representation. 
Centre: Simplified annular geometry. Right: Strip geometry 
obtained via the mapping $w=\ln z$. The defect line is always shown dashed. 
From \cite{Eisler/Peschel10}. Copyright Wiley-VCH,
reprinted with permisson.}
\label{fig:confgeom}
\end{figure}

The variation of $c_\mathrm{eff}$ with $t$ can be determined numerically, but it
turns out that it can also be calculated analytically \cite{Eisler/Peschel10}. 
Since it is an exercise in going to two dimensions and using partition functions 
as in section 3, it is presented here briefly. Because one is at the critical point, 
one can use conformal mappings. The scheme is shown in fig. \ref{fig:confgeom}.

%
\begin{figure}
\center
\includegraphics[scale=0.7]{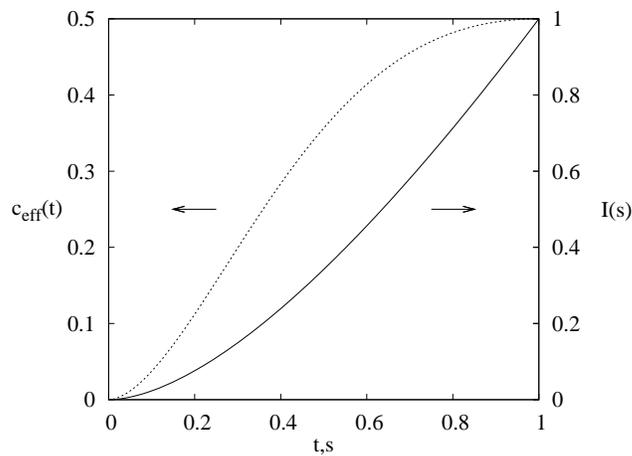}
\caption{Effective central charge $c_\mathrm{eff}(t)$ for a TI chain as a function of the 
defect strength $t$ (left curve). From \cite{Eisler/Peschel10}. Copyright Wiley-VCH,
reprinted with permisson.}
\label{fig:ints}
\end{figure}

%
In the end, one obtains an expression for the $\varepsilon_l$ with a gap which one can 
insert into the continuum formula (\ref{eqn:Scrit1}). The integrals lead to dilogarithms
in terms of a parameter $s=2/(t+1/t)$ which is the transmission amplitude 
through the defect, i.e. $s^2$ is the transmission coefficient. The formula is somewhat
long, so it is more instructive to show the result graphically, see fig. \ref{fig:ints}.
For the R\'enyi entropy $S_2$, by the way, one finds a very simple result, namely
\begin{equation}
c_\mathrm{eff,2} = \frac {8}{\pi^2} \arcsin^2(s/\sqrt2)
   \label{defect_Renyi}
\end{equation}
%

%
%

The continuous variation of the coefficient might seem natural, but it is connected
with the free-fermion nature of the TI and the hopping chain. The defect is then a
``marginal'' perturbation which changes also the local magnetic exponent continuously.
Things are different for a defect in an XXZ chain, which is a Fermi system with interactions.
Then a defect either leads to $c_\mathrm{eff}=0$ if the interaction is repulsive, or is 
irrelevant, i.e. $c_\mathrm{eff}=1$, if the interaction is attractive. This is in analogy 
to the transmission properties in this case.

%
%

\subsection{\bf{Inhomogeneous systems}}

The entanglement can decrease or increase if one makes a system inhomogeneous. This is illustrated
here with two simple but instructive examples.
\vspace{0.2cm}

\noindent (a) {\it{Hopping chain in a field}} \cite{Eisler/Igloi/Peschel09}

Consider an open chain of $2L$ sites with Hamiltonian
\begin{equation} 
H=- \frac {1}{2} \sum_{n=-L+1}^{L-1} (c_n^{\dagger} c_{n+1}+ c_{n+1}^{\dagger} c_n) +
    h \sum_{n=-L+1}^{L}(n-1/2)c_n^{\dagger} c_n
\label{hopping_gradient}
\end{equation}
This describes the so-called Wannier-Stark problem of electrons in a constant electric field.
In magnetic language, it is an XX chain with a linearly varying magnetic field in the $z$-direction.
Due to the field, the particles accumulate on the left.\\

\noindent \emph{Features}
\begin{itemize}
\item Density profile, system full on the left and empty on the right
\item Characteristic length $\lambda=1/h$
\item Transition region has width $2\lambda$
\item Single-particle wave functions are Bessel functions $\phi_k(n)=J_{n-k}(1/h)$, \\concentrated near 
      site $k$.
\item Single-particle energies are equidistant, $\omega_k=h(k-1/2)$, Wannier-Stark ladder
\end{itemize}
\noindent Correlation matrix for a half-filled system for $L \rightarrow \infty$ 
\begin{eqnarray}
C_{mn}&=& \sum_{k=0}^{\infty} J_{k+m}(\lambda) J_{k+n}(\lambda)\\
      &=&  \frac{\lambda}{2(m-n)} \left[J_{m-1} J_{n} - J_{m} J_{n-1} \right]
\label{eq:corrXXlinf}
\end{eqnarray}
In the limit $\lambda \to \infty$, this reduces to the result (\ref{corrXX}) for the homogeneous chain.

The length scale $\lambda$ is seen also in the low eigenvectors of $\bf{C}$. They are essentially
confined to the transition region.

\begin{figure}[thb]
\center
\includegraphics[scale=.7]{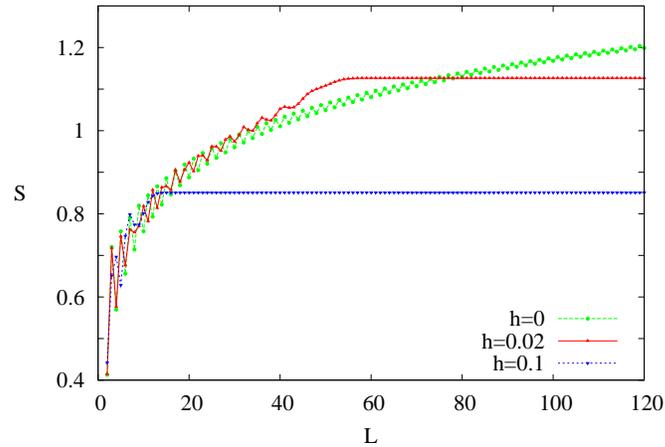}
\caption{Entanglement entropy for a hopping chain in a linear potential as a function
of the half-length $L$. From \cite{Eisler/Igloi/Peschel09}. Copyright IOP Publishing, 
reprinted with permission.}
\label{fig:ent_gradient_XX}
\end{figure}

\noindent Numerical results for the entanglement entropy if the system is divided in the middle are shown 
in fig. \ref{fig:ent_gradient_XX}.  
%

\vspace{0.2cm}
\noindent \emph{Features}

\begin{itemize}
\item Logarithmic up to  $L \approx \lambda$
\item Saturation for $L > \lambda$, if $h \neq 0$
\item Saturation value for large $\lambda$
\begin{equation} 
S_{\infty}(\lambda)=\frac{1}{6}\ln (2\lambda)
\label{eq:entasympt_gradient_XX} 
\end{equation}
\end{itemize}
This is analogous to (\ref{eqn:STI3}), where the correlation length entered. Interpretation:
The parts outside the interface region, which are either full or empty, cannot contribute to the 
entanglement.

\noindent (b) {\it{Inhomogeneous hopping}} \cite{Vitagliano/Riera/Latorre10}

Consider a model with Hamiltonian
\begin{equation} 
H=- \frac {1}{2} \sum_{n=-L+1}^{L-1} t_n (c_n^{\dagger} c_{n+1}+ c_{n+1}^{\dagger} c_n)
\label{hopping_inhom}
\end{equation}
where the hopping amplitudes $t_n$ decay rapidly from the center towards the ends of the chain, 
for example like $t_n=\exp(-|n|)$. In this model, the density in the ground state is constant as for
a homogeneous chain. However, the state is \emph{highly entangled}.\\

\noindent \emph{Example:} Four sites

\begin{figure}[thb]
\center
\includegraphics[scale=.35]{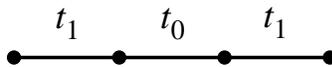}
\caption{Four-site chain with corresponding hopping amplitudes.}
\label{fig:four-sites}
\end{figure}
%

For $t_1 \ll t_0$, the lowest single-particle states have 
energies $\omega_1=-t_0$ and $\omega_2=-t_1^2/t_0$. These states are occupied in the ground state 
and the corresponding eigenvectors are approximately

\vspace{0.1cm}
\begin{equation}
 \phi_1= \frac {1}{\sqrt2} 
     \left(  \begin{array} {ll}
      0 \\  
      1 \\
      1 \\
      0    
  \end{array}  \right), \quad\,\quad
 \phi_2= \frac {1}{\sqrt2}  
     \left(  \begin{array} {rr}
      1 \\  
      0 \\
      0 \\
     -1    
  \end{array}  \right), 
 \label{matrixA}
\end{equation}
\vspace{0.1cm}

\noindent In the first one, sites 2 and 3 are fully entangled, in the second one 
sites 1 and 4. 

\pagebreak

\noindent The total correlation matrix is

\begin{figure}[thb]
\center
\includegraphics[scale=.35]{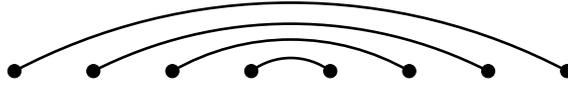}
\caption{Concentric entanglement structure in an inhomogeneous hopping model. After Vitagliano
         et al. \cite{Vitagliano/Riera/Latorre10}.}
\label{fig:vitagliano}
\end{figure}

\vspace{0.1cm}
\begin{equation}
 {\bf{C}} = \frac{1}{2}  
    \left(  \begin{array} {rrrr}
      1 & 0 & 0 & -1 \\  
      0 & 1 & 1 & 0  \\
      0 & 1 & 1 & 0 \\
      -1 & 0 & 0 & 1
  \end{array}  \right)   
  \label{Cinhom}
\end{equation}
\vspace{0.2cm}
Restricting {\bf{C}} to the left or right half-chain, one finds
$\zeta_1=\zeta_2=1/2$, i.e. $\varepsilon_1= \varepsilon_2=0$, which gives $S=2\ln 2$.  The mechanism
persists for larger systems and leads to the concentric structure shown in fig. \ref{fig:vitagliano}.

%
%
\subsection{\bf{Entropy and fluctuations}}

In hopping models, there is a close connection between the entanglement entropy and
the particle-number fluctuations in the considered subsystem. This allows to put 
bounds on $S$ \cite{Wolf06,Gioev/Klich06}.

In terms of the eigenvalues $\zeta_l$ of the correlation matrix {$\bf{C}$}, one has
\begin{equation} 
S= -\sum_l \left[ \zeta_l \ln \zeta_l + (1-\zeta_l)\ln (1-\zeta_l) \right] = \sum_l s( \zeta_l)
   \label{Szeta}
\end{equation}
The function $s(x)$ defined by minus the bracket in (\ref{Szeta}) has the properties
\begin{itemize}
\item Symmetry with respect to $x=1/2$
\item $s(x)=0$ for $x=0$ and $x=1$
\item Maximum at $x=1/2$ with $s(1/2)= \ln 2$
\end{itemize}

\begin{figure}[thb]
\vspace{0.3cm}
\center
\includegraphics[scale=.25]{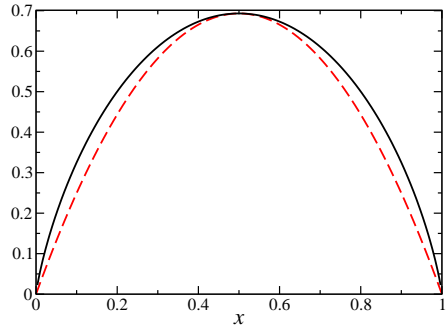}
\caption{The function $s(x)$ (solid) and its quadratic lower bound (dashed).}
\label{fig:bound}
\end{figure}

\noindent As a result, it can be bounded in $0 \le x \le 1$ by a parabola

\begin{equation}
s(x) \ge 4 \ln 2 \; x(1-x)
   \label{inequality1}
\end{equation}
and the equality holds for $x=0,1/2,1$. This is shown graphically in fig. \ref{fig:bound}
\vspace{0.3cm}

%

It follows that
\begin{equation}
S \ge 4 \ln 2 \,\sum_l \zeta_l(1-\zeta_l) = 4 \ln 2 \,\,\mathrm{tr} \bigl [\bf{C}(1-\bf{C})\bigr ]
   \label{inequality2}
\end{equation}
But the traces can be written as 
\begin{equation}
\mathrm{tr} {\bigl [\bf{C}(1-\bf{C})\bigr ]} = \langle \,N^2 \,\rangle - \langle \,N \,\rangle^2
   \label{numbers}
\end{equation}
where $N=\sum_i c_i^{\dagger}c_i$ is the particle number operator in the subsystem.
Therefore the particle-number fluctuations give a lower bound on $S$
\begin{equation}
S \ge 4 \ln 2 \,\,\bigl [ \langle \,N^2 \,\rangle - \langle \,N \,\rangle^2 \,\bigr ]
   \label{bound}
\end{equation}
These fluctuations have a direct physical significance and are easier to calculate.

\vspace{0.2cm}
\noindent \emph{Application}
\begin{itemize}
\item One dimension, large $L$
\begin{equation}
      \bigl [ \langle \,N^2 \,\rangle - \langle \,N \,\rangle^2 \,\bigr ]= 
       \frac {1}{\pi^2} \ln L
   \label{fluct1}
\end{equation}
\item Two dimensions, large $L$ 
\begin{equation}
      \bigl [ \langle \,N^2 \,\rangle - \langle \,N \,\rangle^2 \,\bigr ] \sim
       L \,\ln L
   \label{fluct1}
\end{equation}
\end{itemize}
By shifting the parabola $x(1-x)$ upwards, one can also obtain upper bounds. In this way, one can 
\emph{prove} the behaviour of the entropy in various dimensions without actually calculating
it. The lower bound in 1D gives the prefactor $4\ln 2/\pi^2=0.28$, which is rather close to
the exact value $1/3$. From these considerations, a general formula for the prefactor was 
obtained which involves an integral over the surface of the subsystem in real space and the
Fermi surface in momentum space \cite{Gioev/Klich06}.

\pagebreak


\section{Quenches and miscellaneous}

So far we have been concerned with time-independent situations. In this last section,
we turn to cases where the entanglement changes in time. Moreover, I return once
more to possible relations between the entanglement Hamiltonian and the real one
and finally give a short summary.  
%
%

\subsection{\bf{Quenches}}

If a quantum state changes in time, this will in general affect the entanglement
properties. However, the change must be more than a mere phase factor. Thus
one has to have a time evolution with a Hamiltonian, for which $|\Psi\rangle$ is
not an eigenstate. The simplest set-up is to make an instantaneous change
\begin{equation} 
H_0 \rightarrow H_1
   \label{def_quench}
\end{equation}

\noindent After that
\begin{itemize}
\item The state $|\Psi\rangle$ evolves as $|\Psi(t)\rangle = e^{-iH_1t}|\Psi_0\rangle$.
\item The total density matrix $\rho$  evolves.
\item The RDM's $\rho_{\alpha}$ also evolve.
\end{itemize}

If $H_1$ is a free-particle operator, the arguments work as before. If the initial
state was a Slater determinant, the correlation functions at time $t$ 
\begin{equation} 
\langle\Psi(t)| c_m^{\dagger}c_n^{\dagger}c_kc_l|\Psi(t)\rangle=
\langle\Psi_0| c_m^{\dagger}(t)c_n^{\dagger}(t)c_k(t)c_l(t)|\Psi_0\rangle
   \label{corr_equal}
\end{equation}
factor again, because the Heisenberg operators $c_k(t)$ at time $t$ are then linear 
combinations of the initial ones. Therefore $\rho_{\alpha}(t)$ has the exponential form 
(\ref{rhogen}) but with a time-dependent operator $\mathcal{H}_\alpha(t)$ and the eigenvalues 
$\varepsilon_l(t)$ follow from the correlation matrix at time $t$ 
\begin{equation}
C_{i,j}(t)=
\langle \Psi_0| \, c_i^{\dagger}(t) \, c_j(t) \, |\Psi_0 \rangle \, .
\label{eq:corrt}
\end{equation}
Therefore, one only needs to determine the time evolution of the operators
$c_j(t)$ in the Heisenberg picture. 

\noindent Physically, one finds a surprising phenomenon, namely the entanglement
\emph{increases} after the quench

\begin{itemize}
\item In global quenches $S \sim t$ 
\item In local quenches, $S \sim \ln t$
\end{itemize}
We show this explicitly for two examples.
%
%

\subsection{\bf{Global quench}}

\vspace{0.2cm}
\noindent Hopping model

\begin{itemize}
\item Start from fully dimerized, half-filled model, only pairs of sites $(2n,2n+1)$ 
      are coupled and correlated.
\item Make it homogeneous with dispersion relation $\omega_q=-\cos q$ and let it evolve. 
\end{itemize}
The time evolution of the Fermi operators then involves Bessel functions
\begin{equation}
c_j(t)=\sum_m i^{j-m}J_{j-m}(t)c_m
\label{eq:bessel}
\end{equation}
and the result for the correlation matrix is
\eq{\fl\hspace{1.5cm}
C_{m,n}(t) = \frac {1}{2}  \left[\delta_{m,n} + \frac {1}{2}(\delta_{n,m+1}+
\delta_{n,m-1}) +  e^{-i \frac {\pi}{2}(m+n)} \frac {i(m-n)}{2t}J_{m-n}(2t)
\right] \label{eq:corrt_global}
}
%
The resulting single-particle spectra are shown on the left of Fig. 
\ref{fig:globalquench}.
%
\begin{figure}[thb]
\center
\includegraphics[scale=0.55]{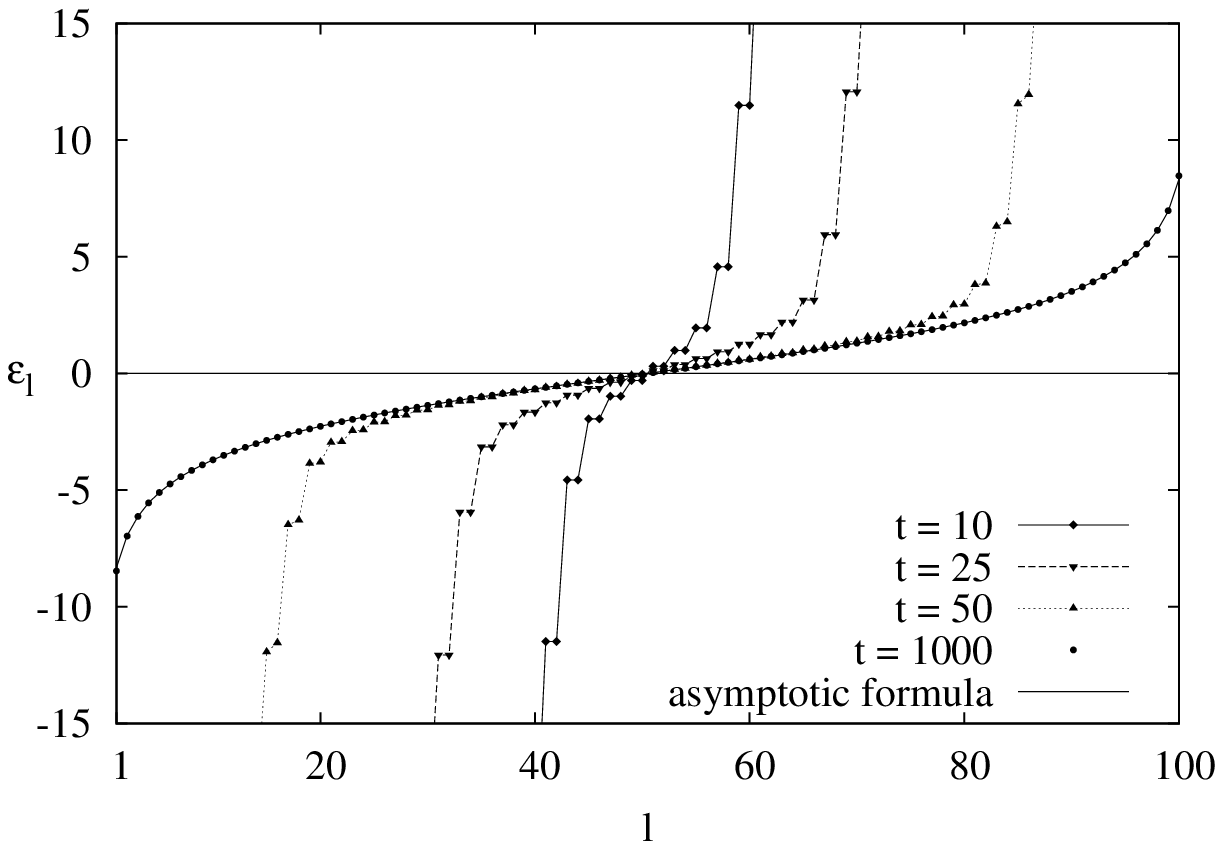}
\includegraphics[scale=0.55]{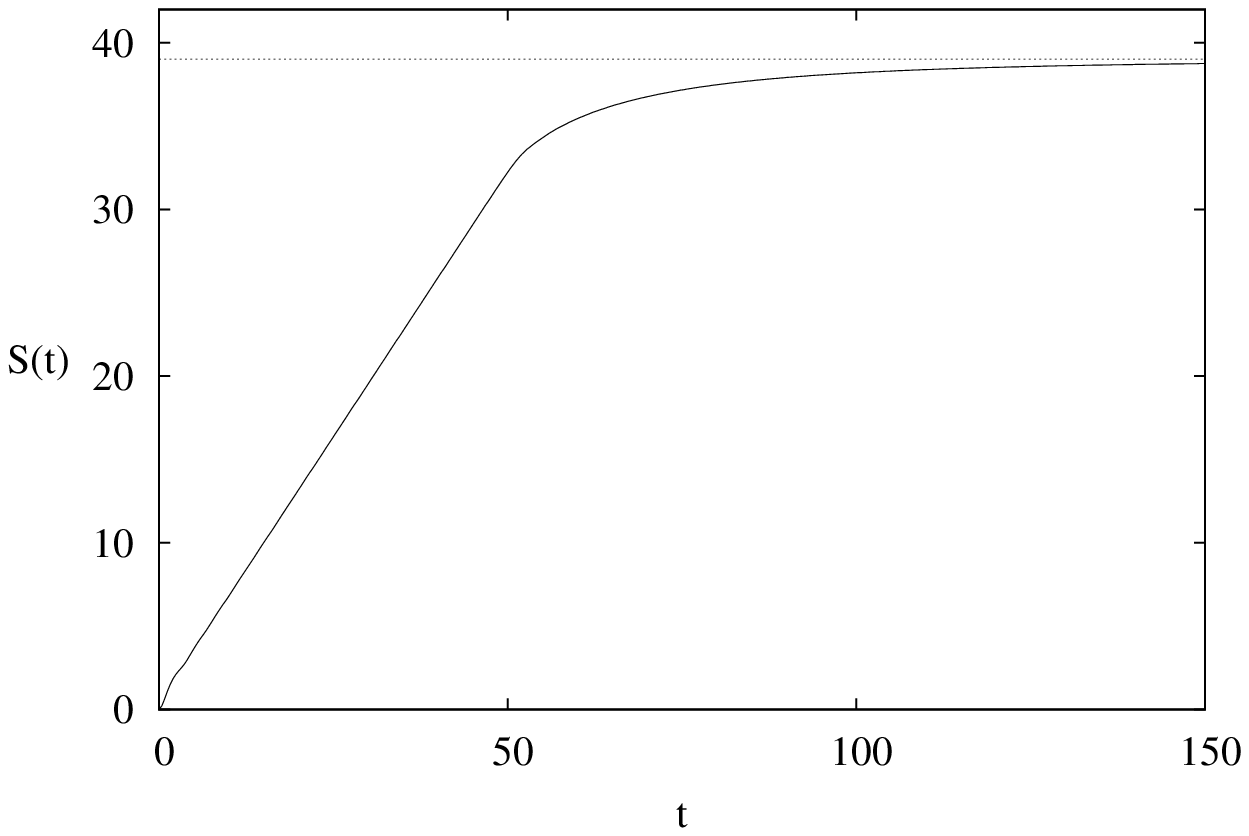}
\caption{Global quench in a hopping model, starting with a fully dimerized initial state.
Left: Time evolution of the single-particle spectrum for a segment of $L=100$ sites. 
Right: Entanglement entropy with the asymptotic value. From \cite{Eisler/Igloi/Peschel09}.
Copyright IOP Publishing, reprinted with permission.}
\label{fig:globalquench}
\end{figure}
%

\noindent \emph{Features}
\begin{itemize}
\item Dispersion linear near zero
\item Slope decreases with time, $S$ increases
\item For times $t \gg L/2$ approach to a limiting curve, $S$ saturates
\end{itemize}
The asymptotic form of the spectrum follows from the first three terms in 
(\ref{eq:corrt_global}) which correspond to a tridiagonal correlation matrix 
and are the Fourier transform of the constants $\langle c_q^{\dagger}c_q \rangle$ 
in the initial state. The eigenvalues for a segment are
\eq{
\zeta_l(\infty)= \frac {1}{2} (1+\cos q_l), \;\;\; q_l =\frac {\pi}{L+1}l,
 \;\;\; l=1,2...L
\label{eq:zetainf}
}
and lead to
\eq{\varepsilon_l(\infty)= 2 \ln \tan(q_l/2).
\label{eq:epsinf}}
The spacing of the $q_l$ is proportional to $1/L$ and gives an \emph{extensive} 
entropy $S=L(2\ln 2 - 1)$.

\vspace{0.2cm}
\noindent The build-up of an extensive entropy is a typical signature of global quenches.
 
\begin{figure}[thb]
\center
\includegraphics[scale=0.45]{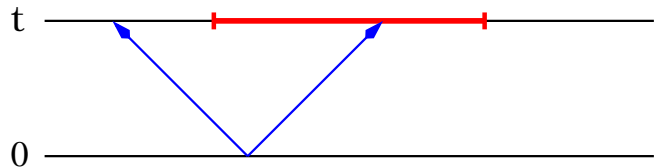}
\caption{Creation of entanglement after a global quench by emitted particle-hole pairs for 
the case of a segment in a chain.}
\label{fig:light-cone}
\end{figure}

\noindent It has a simple physical interpretation due to Calabrese and Cardy \cite{CC05} 
sketched in fig. \ref{fig:light-cone}.
\begin{itemize}
\item Particle-hole pairs are emitted
\item Create entanglement between the subsystem and remainder
\item Travel with maximum velocity $v=1$
\item ``Light-cone effect'', $S \sim t$ as long as separation $2t < L$
\end{itemize}
%
%
%
%
The result is relevant for numerical calculations, because it means that one can
follow the evolution only for a limited time with DMRG. Beyond that, the state
is too entangled to be well approximated.

%
%

\subsection{\bf{Local quench}}
\begin{figure}[thb]
\center
\includegraphics[scale=0.55]{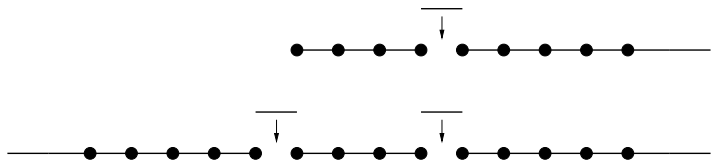}
\caption{Two variants of a local quench.}
\label{fig:geolocalquench}
\end{figure}

\noindent Hopping model, set-up shown in fig. \ref{fig:geolocalquench}

%
%

\begin{itemize}
\item Initially subsystem (center) decoupled from the rest 
\item Add bond(s) to create a homogeneous chain and let system evolve
\end{itemize}

\begin{figure}[thb]
\center
\includegraphics[scale=0.58]{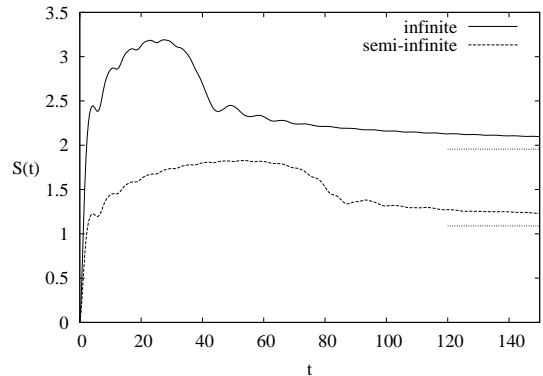}
\caption{Entanglement entropy for the two geometries after the quench for a subsystem of
   length $L=40$. From \cite{EKPP08}. Copyright IOP Publishing, reprinted with permission.}
\label{fig:entropylocal}
\end{figure}

The evolution of the Fermi operators is again given by (\ref{eq:bessel}), but the initial
condition is different. The calculation has to be done numerically.
In fig. \ref{fig:entropylocal} the result for $S$ is shown.
%
%
\pagebreak

\begin{figure}[thb]
\center
\includegraphics[scale=0.55]{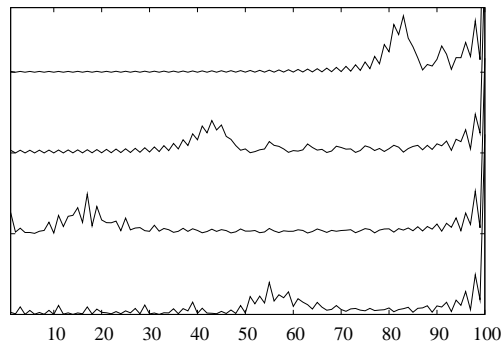}
\caption{Front propagation in the lowest single-particle eigenvector for the semi-infinite
geometry and $L=100$. Shown are the times $t=20,60,120,160$. From \cite{EKPP08}. 
Copyright IOP Publishing, reprinted with permission.}
\label{fig:frontsemi}
\end{figure}

\noindent \emph{Features}
\begin{itemize}
\item ``Entanglement bursts'' after the connection
\item Duration $t=L$ (infinite case) and $t=2L$ (semi-infinite case)
\item For larger times approach to equilibrium (dotted)
\end{itemize}
The plateau can be related to a front which starts from the initial defect site and 
travels through the subsystem until it leaves it again. This is seen directly in the
lowest eigenvector in fig. \ref{fig:frontsemi}.  
%
%
Using methods of conformal field theory, one can derive analytical formulae for 
both cases \cite{CC07,EKPP08}
\eq{S(t)= \nu \frac c 6 \ln\left[ \frac{4L}{\nu \pi}t
\sin \left(\frac{\nu \pi t}{2L} \right) \right] + k_{\nu}
\label{eq:ent_cft}}
where $\nu$ is the number of contact points and $k_{\nu}$ is a constant which depends 
on the geometry. This formula is in good agreement with the numerical data. 
For $t \ll L$, it gives a \emph{logarithmic} entropy growth. If $L \rightarrow \infty$ this
persists for \emph{all} times.

For numerical calculations, this is a more favourable situation. One can follow the
evolution a much longer time.

%
\subsection{\bf{Periodic switching}}
\begin{figure}[thb]
\center
\includegraphics[scale=0.55]{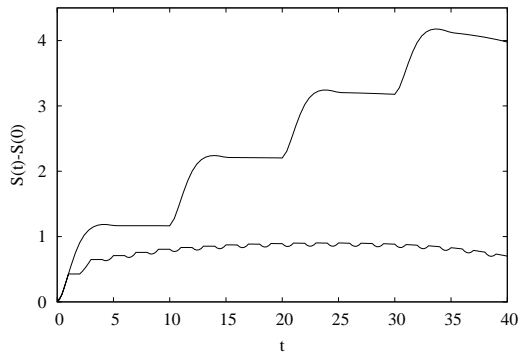}
\caption{Entropy evolution for periodically connected chains and $L=40$.
         Upper curve: $\tau=5$, lower curve: $\tau=1$. From \cite{review09}. 
         Copyright IOP publishing, reprinted with permission.}
\label{fig:entro_perq}
\end{figure}

An interesting effect appears if one connects and disconnects two half-chains periodically
for a certain time $\tau$. One can call this a periodic local quench. Numerical results
for the entanglement are shown in fig. \ref{fig:entro_perq}.\\ 

%
%
\noindent \emph{Features}
\begin{itemize}
\item Switching directly visible
\item Rapid switching: logarithmic increase 
\item Slow switching: linear increase
\end{itemize}
The curve for rapid switching resembles the result for a single quench, compare 
fig. \ref{fig:entropylocal}. This can be understood as follows. The time-evolution 
operator for one period is
\eq{
 U = U_0 \, U_1 = \ee^{-i H_0 \tau} \ee^{-i H_1 \tau}
\label{eq:u_perq}}
where $H_0$ and $H_1$ are the Hamiltonians for the the two configurations and do not
commute. However, for small $\tau$, one can take the same Hamiltonian limit as for
the transfer matrices in section 3 and combine the exponentials. Then
\eq{
U= \ee^{-i \bar{H} 2\tau} \quad , \quad \bar{H} = \frac {1}{2}(H_0 + H_1)
\label{eq:u_heff}}
The average time evolution therefore corresponds to a \emph{single} local quench 
where the final system has a defect with reduced hopping amplitude $t'=t/2$ at the contact. 
For such a case, the evolution of $S$ is similar as for a quench to a homogeneous system
and the behavour is logarithmic in time.

The curve for slow switching rises on average \emph{linearly}. The interpretation is that 
here the disconnected system has enough time to ``recover'' and thereby the entanglement 
gain repeats itself after each new connection. The problem can be treated analytically in 
a continuum model \cite{Klich/Levitov09}.

In general, one can express $S$ in terms of the (time-dependent) cumulants of the 
probability distribution $P_n$ to transfer $n$ particles. This provides a link to
the so-called ``full counting statistics'' of the junction and thus in principle to
measurable quantities. For the example given here, the distribution is Gaussian and
only the second cumulant enters.

%
\subsection{\bf{Entanglement Hamiltonian and subsystem Hamiltonian}}
The thermal form of the RDM automatically leads to the question whether there is a relation
between $\mathcal{H}_{\alpha}$ and $H_{\alpha}$. In section 2.6 we have already seen that 
in general this is not so. But are there cases, where a relation exists ? 

The answer is yes. For example, it has been seen in Heisenberg ladders, where the subsystem 
was chosen as one of the two legs. We discuss here an example, which is somewhat simpler
and a free-fermion model \cite{Peschel/Chung11}.\\

%
\begin{figure}[thb]
\center
\includegraphics[scale=0.45]{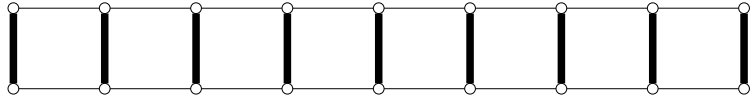}
\caption{Ladder geometry for a fermionic hopping model. \\ The subsystem is chosen as one 
of the legs.}
\label{fig:globalquench}
\end{figure}
%

Consider a hopping model on a ladder with opposite dispersion in both legs and hopping with
amplitude $\delta$ between them. The Hamiltonian is
\begin{equation}
H = H_1 + H_2 + H' = \sum_q \gamma_q\, a_q^{\dag}a_q - \sum_q \gamma_q\,b_q^{\dag}b_q + 
                      \sum_q \delta\, (a_q^{\dag}b_q+b_q^{\dag}a_q)   
\label{ladder}
\end{equation}
Diagonalizing (\ref{ladder}) with a canonical 
transformation 
\begin{equation}
a_q=u_q\alpha_q+v_q\beta_q, \quad b_q=-v_q\alpha_q+u_q\beta_q, \quad u_q^2+v_q^2=1,
\label{ladder_trans}
\end{equation}
one obtains
\eq{
H =  \sum_q \omega_q (\alpha_q^{\dag}\alpha_q - \beta_q^{\dag} \beta_q) \,\, ,\hspace{1cm}
        \omega_q= \sqrt{\gamma_q^2+\delta^2}
\label{ladder2}}
From that, one obtains the correlation matrix. Due to the translation invariance, it
is diagonal in momentum space and in the subsystem 1 of the $a's$ one has   
\eq{
\zeta_q= <a_q^{\dag}a_q>=v_q^2 = \frac {1}{2}(1- \frac{\gamma_q}{\omega_q}) 
\label{occupation}}
This gives the single-particle eigenvalues
\eq{
\varepsilon_q =\ln \left(\frac{\omega_q+\gamma_q}{\omega_q-\gamma_q}\right)
\label{epsilon_ladder}}
and $\mathcal{H}_{1}$ has the form
\eq{
\mathcal{H}_{1} =  \sum_q \varepsilon_q a_q^{\dag}a_q
\label{ent_ladder}}
If now the rung hopping $\delta$ is large, one obtains $\varepsilon_q= 2\gamma_q/\delta$ 
and the relation
\eq{
 \mathcal{H}_1 = \frac {2}{\delta} H_1
\label{relation}}
This is a direct proportionality between the two Hamiltionians. 

\vspace{0.2cm}
\noindent \emph{Remarks} 

\begin{itemize}
\item Holds for dominating rung couplings
\item Follows from first-order perturbation theory in $H_1+H_2$ 
\item Entanglement near maximum $S=L \ln 2$
\item Entropy extensive due to long interface
\end{itemize}

For arbitrary $\delta$, the single-particle energies $\varepsilon_q$ and $\gamma_q$ are
not proportional to each other. Therefore the hopping range in $\mathcal{H}_1$ is in 
general different from that in $H_1$.

%
%

\subsection{\bf{Concluding remarks}}

I have given an account of the entanglement properties of solvable models,
either free particle or integrable, and shown in particular that 

\begin{itemize}
\item One is lead to a thermodynamic problem
\item A particular Hamiltonian enters 
\item Its spectrum determines the Schmidt weights
\item The ground states of homogeneous chains are weakly entangled
\item Global quenches lead to strongly entangled states 
\end{itemize}

Almost all considerations had to do with lattice models. These are the systems
one studies in numerical investigations motivated by solid state physics or uses 
in quantum information. They also have the advantage that no divergencies appear in
finite geometries.

This does not mean that continuum systems are unimportant. The first calculations 
of entanglement entropies took place in the context of black-hole theory 
and thus in a continuum setting. And the use of conformal invariance has not
only shown a deeper connection between the various models but also allowed to
derive many special results. But that would be a lecture series in its own.
Those who are interested can find a lot of material in a special issue of
J. Phys. A 42 (2009). There, entanglement for free quantum fields is reviewed by 
Casini and Huerta \cite{Casini/Huerta09} and within conformal field theory by 
Calabrese and Cardy \cite{CC09}.


\pagebreak

\noindent {\bf{Acknowledgement}}

I would like to thank Francisco Alcaraz for the invitation to lecture at the school and 
the International Institute of Physics for its financial support and the hospitality
at Natal. I also thank Ming-Chiang Chung, Viktor Eisler and Jos\'e Hoyos for a substantial
number of figures.

\section*{References}

\providecommand{\newblock}{}

\end{document}